\documentclass[preprint,3p,10pt]{elsarticle}

\makeatletter
\def\ps@pprintTitle{%
	\let\@oddhead\@empty
	\let\@evenhead\@empty
	\let\@evenfoot\@oddfoot}
\makeatother

\usepackage{lineno}

\usepackage{geometry}

\usepackage{graphicx,subfigure}
\DeclareMathAlphabet{\mathbmit}{OML}{cmm}{b}{it}

\usepackage{siunitx}
\DeclareUnicodeCharacter{00A0}{ }
\usepackage{amsmath}
\usepackage{caption}
\usepackage{booktabs}
\usepackage{mathtools}
\usepackage{siunitx}  
\usepackage{tikz} 
\usepackage{pgfkeys}
\usetikzlibrary{shapes, arrows.meta, bending, calc, fit, matrix, decorations.markings, positioning, plotmarks}

\usepackage{xcolor}
\colorlet{color}{blue}

\usepackage[normalem]{ulem}

\biboptions{sort&compress}

\usepackage{geometry}
\usepackage{tikz}

\makeatletter
\renewenvironment{abstract}
  {\global\setbox\absbox=\vbox\bgroup            
   \hsize=\textwidth
   \noindent
   \ignorespaces}%
  {\egroup}
\makeatother

\begin{document}

\begin{frontmatter}

\title{A compact quasi-zero stiffness metamaterial based on monolithic shells for vibration isolation}


\author[add1]{Yong Zhang}
\author[add1,add3,ca]{Xianfeng Chen}

\address[add1]{Department of Precision and Microsystems Engineering, Delft University of Technology, Mekelweg 2, 2628 CD, Delft, The Netherlands}
\address[add3]{A*STAR Quantum Innovation Centre(Q.InC), Institute for Materials Research and Engineering(IMRE), Agency for Science, Technology and Research(A*STAR), 2 Fusionopolis Way, 08-03 Innovis 138634, Singapore}

\address[ca]{Corresponding author email: {xianfeng\_chen@imre.a-star.edu.sg}}

\begin{abstract}
Quasi-zero stiffness (QZS) metamaterials are highly effective in isolating objects from low-frequency external vibrations, due to their high static stiffness but low dynamic stiffness characteristics. Traditionally, QZS metamaterials are designed by combining a negative-stiffness part with a positive-stiffness counterpart. Here, we present a novel QZS metamaterial design without relying on combining two components. The QZS characteristic is achieved solely through monolithic shell elements' unique geometry and nonlinear deformation. Using experimental and numerical approaches, we investigate the static and dynamic responses of the proposed metamaterials as a function of their geometric parameters. We then tune the structure's geometry to achieve ideal zero-stiffness behaviors and experimentally demonstrate an exceptional low-frequency vibration isolation mechanism. This concept can be further utilized as a building block for constructing metamaterials with multiple zero-stiffness features, enabling a broad range of applications.

\end{abstract}

\end{frontmatter}


\graphicspath{{figures/}}

\section*{Introduction}
Vibration isolation plays a crucial role in a wide range of engineering applications and fundamental research, such as high-precision motion platforms \cite{harris2002harris,heertjes2020control,yoon2010analysis}, aerospace systems \cite{xu2016intelligent, liu2015recent}, civil structures \cite{chapain2019vibration,ismail2015inner}, and ultra-precision sensing \cite{accadia2011seismic,braccini2005measurement,matichard2015seismic}. Using vibration isolation, the dynamic performance of the systems can be significantly improved without suffering from external vibration disturbances. Nowadays, there are mainly two types of vibration isolation strategies, namely active and passive isolation. Active isolation systems allow real-time control of stiffness and force, but normally require a series of actuators and controllers, resulting in a costly and complex design of mechatronic systems \cite{yun2011general, xianmin2002active}. On the other hand, passive vibration isolation is realized by using proper isolation mechanisms without the use of any control unit. To achieve effective isolation, the natural frequency of these isolation mechanisms should be designed to be as low as possible, since the effective frequency range for isolation typically starts at $\sqrt{2}$ times the natural frequency \cite{li2020negative}. In addition, it is essential to ensure that these mechanisms can maintain sufficient static stiffness to support the load of the object to be isolated. Traditional linear vibration dampers with constant stiffness \cite{narimani2004frequency} cannot simultaneously meet the requirements for both stiffness and frequency. To overcome this limitation, quasi-zero stiffness (QZS) mechanisms have been proposed and actively investigated in the field of passive vibration isolation \cite{yang2021novel,liu2024quasi}. 

QZS mechanisms refer to structures that possess high static stiffness in their undeformed configurations while exhibiting a dynamic stiffness close to zero under a payload. Unlike linear structures, QZS mechanisms typically undergo large deformations, resulting in distinct stiffness in their deformed and undeformed configurations. Conventionally, this stiffness characteristic is realized by combining positive and negative stiffness elements in parallel, such that the total stiffness approaches zero within a certain displacement range. Based on this principle, some QZS mechanisms have been proposed, mainly using springs \cite{jing2022situ,carrella2007static,carrella2009force, huang2014vibration}, magnetic elements \cite{robertson2009theoretical,jiang2020design,carrella2007optimization,yan2019nonlinear,yan2018vari,yan2022lever,liu2024nonlinear}, and mechanism-based designs \cite{yan2022bio,jing2019novel,wu2020mechanical}. For instance, Carrella \textit{et al.} have proposed a type of spring construction and derived the transmissibility of such QZS vibration isolators. Ishida \textit{et al.} \cite{ishida2017design} constructed a zero-stiffness prototype using a combination of oblique springs with rigid constraints. Yan \textit{et al.}  \cite{yan2020large} have proposed a three-link mechanism design to increase the vibration stroke. However, these rigid mechanical constructions are typically used for large-scale systems and are less ideal for small-scale components because of the need to combine multiple parts. 


The rapid development of mechanical metamaterials provides a new option for designing QZS isolators. Mechanical metamaterials are rationally designed periodic structures engineered to exhibit unconventional properties, such as negative Poisson's ratio \cite{bertoldi2017flexible,babaee20133d,huang2016negative}, multi-stability \cite{zhang2021novel,xiu2022topological,zhang2020design,mofatteh2022programming}, and negative thermal expansion \cite{wu2016isotropic,wang2016lightweight}. For vibration control, many QZS metamaterials have been presented in the literature, for which the non-linear deformation characteristics of elastic elements play an important role \cite{dalela2022review,ji2021vibration,liu2025metamaterial,song2025genetic}. For example, it has been shown that connecting a sinusoidal beam with an arc structure in parallel can achieve zero stiffness and, thus, low transmissibility at low frequencies \cite{banerjee2023simultaneous,dalela2022design,fan2020design}. Similarly, Cai \textit{et al.}\cite{cai2020design} developed a one-dimensional zero-stiffness mechanism based on buckled beams and folded structures. Furthermore, it has been demonstrated that QZS properties can also be realized with torsional modes \cite{zhang2023compliant,yu2022vibration}. However, most QZS metamaterials are still constructed by combining two elements, i.e., negative and positive stiffness parts, which typically results in a complicated design and inevitable mechanical contact. A monolithic design approach can be a preferred solution to overcome this. Currently, only a few monolithic designs have been proposed, mainly limited to beam elements \cite{zhou2021tunable,dalela2024nonlinear,zhang2021tailored,hou2024quasi,liu2024design}. Other forms of monolithic QZS metamaterials, like shells \cite{parisch1978geometrical}, have not been extensively studied, and the associated design requirements are still not fully clear.

To enrich QZS metamaterial design, in this paper, we propose and investigate a novel type of QZS metamaterial {based on doubly-curved shells. This type of shell geometry has not been investigated for vibration isolation, and the associated QZS mechanism as well as the design space are not fully explored}. The proposed design enables zero-stiffness properties without the need for positive-stiffness counterparts. Specifically, this unique zero-stiffness behavior mainly originates from the geometric nonlinearity of the shells. It is found that by tweaking the shell's geometric parameters, QZS can be well preserved over a broad displacement range. Unlike those metamaterials containing two counterparts (positive and negative stiffness), the proposed monolithic design significantly reduces system complexity and thus simplifies design efforts. {Compared to the beam-based and conical shell-based isolators \cite{zhang2021tailored,hou2024quasi,fan2020design,dalela2022design,liu2024compact}, the proposed curved shell is able to realize a high load-bearing capability while still maintaining isolation performance in a low frequency range.} More importantly, these shells can be utilized as building blocks to construct QZS mechanisms with multiple zero-stiffness bandwidths. We demonstrate that by arranging shells in series and in parallel, it is possible to realize programmable vibration isolation properties for various applications.

\noindent

\section*{Results}
\subsection*{\textbf{Design concept}}

\begin{figure}[!ht]
	\centering
	\begin{tikzpicture}
	\node[anchor=south west,inner sep=0pt] (image) at (0,0) {\includegraphics[trim={0cm 0cm 0cm 0cm},clip,width=0.9\textwidth]{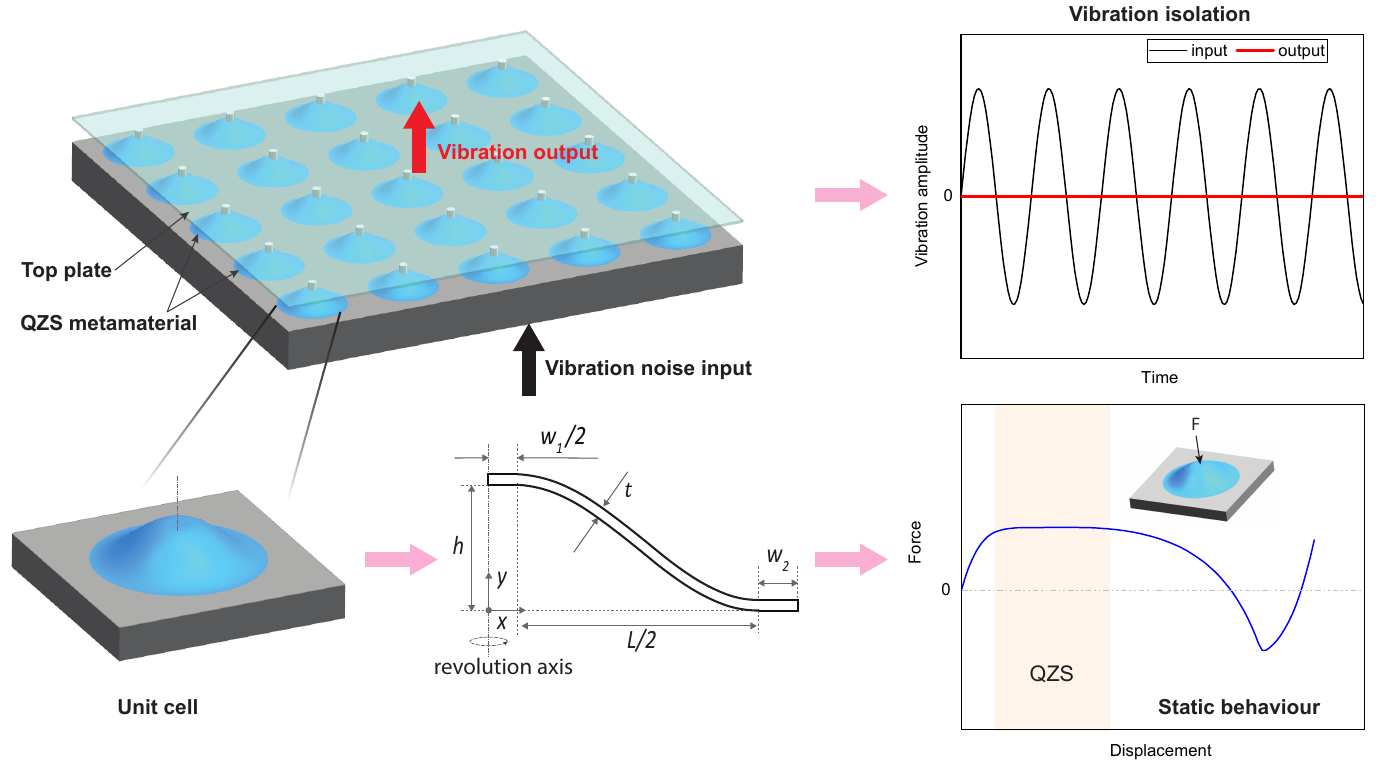}};
	\begin{scope}[x={(image.south east)},y={(image.north west)}]
	\node[] at(0.05,0.9){\small (a)};
         \node[] at(0.05,0.35){\small (b) };
         \node[] at(0.3,0.35){\small (c) };
         \node[] at(0.65,0.9){\small (e)};
         \node[] at(0.65,0.45){\small (d)};
	\end{scope}
	\end{tikzpicture}
	\caption{Schematic of the QZS metamaterial design for vibration isolation. (a) A vibration isolation platform made from metamaterials based on monolithic shells. An object to be isolated will be added to the transparent top plate. (b-c) The functional QZS unit cell is composed of a compliant shell (in blue) and a rigid frame (in grey). The shell is generated by revolving a sinusoidal 2D profile around a revolution axis (y-axis in the sketch), for which the profile is defined by parameters: height ($h$), length ($L$), and thickness ($t$), respectively. {The thickness $t$ is the same across the beam.}  (d) {Simulated} static force-displacement response of a QZS unit cell. When a force is applied at the center, the unit will respond with a constant force plateau at a certain displacement range, i.e., QZS region as highlighted. (e) Illustration of the vibration isolation of the proposed QZS design. When a proper mass is loaded to reach a zero-stiffness plateau, input vibrations will be significantly isolated by the mechanism, leading to a small-amplitude vibration output.}
	\label{structural design}
\end{figure}
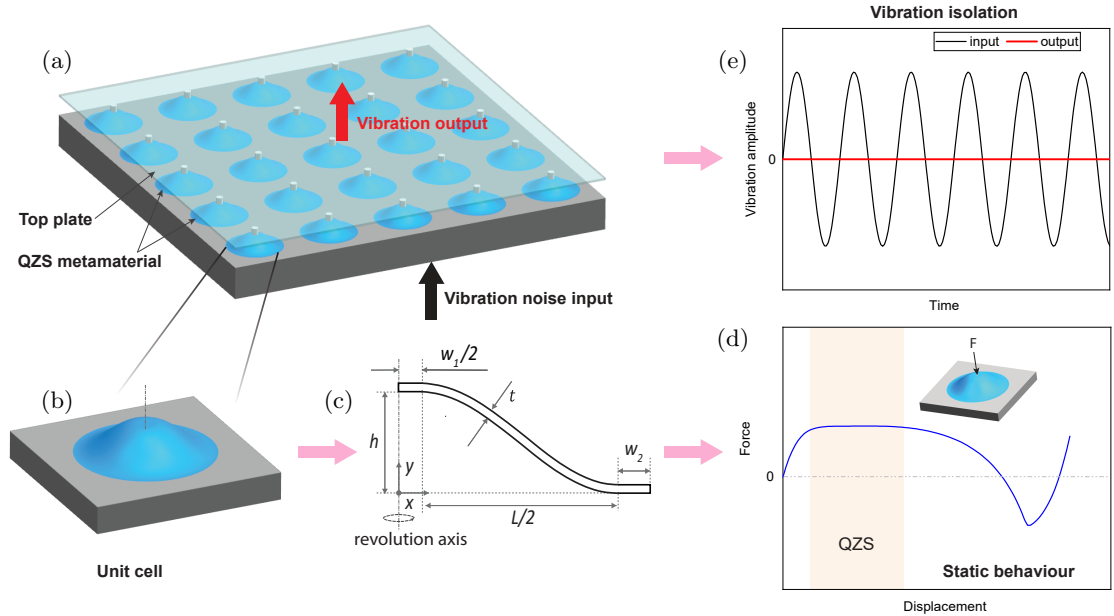

The proposed vibration platform is shown in Fig.~\ref{structural design}(a), which consists of a top plate and a specifically designed metamaterial. The top plate is added for placing the object to be isolated, while the QZS metamaterial is composed of a periodic arrangement of shell-based unit cells. Unlike conventional unit cells originating from a two-dimensional sinusoidal shape profile, we explore the properties of three-dimensional (3D) sinusoidal-shaped shells, which demonstrate different stiffness characteristics due to their unique structural geometry and boundary constraints. In particular, by revolving a sinusoidal-shaped profile around a rotational axis, a 3D shell can be constructed. As illustrated in Fig.~\ref {structural design}(c), the geometry of the shell can be described with the parameters of height ($h$), thickness ($t$), and length ($L$), respectively. The flat regions denoted by $w_1$ and $w_2$ are designed for applying loads such that the shell can be connected to a rigid frame to form a functional unit cell (see Fig.~\ref{structural design}(b)). The proposed functional unit forms a basis for constructing metamaterials. Note that the design parameters of this study refer to $h$, $t$, and $L$, which play an important role in the quasi-zero-stiffness behavior. Specifically, this sinusoidal profile can be expressed using the following Eq.~\eqref{shape}, where the $x-y$ coordinate system is depicted in Fig.~\ref {structural design}(c):
\begin{equation}
{y} =\frac{h}{2} \left( 1 + \cos({\pi} \frac{2x - w_1}{L})\right) \quad \textrm{for} \quad x \in [\frac{w_1}{2}, \frac{L+w_1}{2}]. \label{shape}
\end{equation}

The proposed unit cell design plays a key role in realizing quasi-zero stiffness, as demonstrated by its static force-displacement ($F-d$) response shown in Fig.~\ref {structural design}(d). When a vertical force ($F$) is applied at the center, a positive tangent stiffness is first observed. As the force increases, the stiffness decreases to near zero. Thus, the force-displacement curve shows a plateau with a constant force, i.e., QZS behavior, as highlighted in Fig.~\ref {structural design}(d) (in light peach). Note that the initial positive stiffness allows it to maintain its load-bearing capability for sustaining payloads. These QZS stiffness characteristics make the proposed unit ideal for vibration isolation applications. When a proper object is placed on top of the shell unit, the payload compresses the shell into the QZS region, leading to a very low resonance frequency. From a dynamics perspective, the low-frequency characteristics enable the unit to effectively isolate vibration disturbances (input) and thus result in a low amplitude vibration (output), as depicted in Fig.~\ref {structural design}(e). 

Different from most zero-stiffness metamaterials containing positive and negative stiffness parts, the QZS property here is mainly attributed to the shell's inherent nonlinear large-deformation characteristics. This eliminates the need to connect multiple parts, significantly reducing design complexity. The structural simplicity also allows for integrating a series of functional units into a metamaterial array. By appropriately arranging these units, the resulting QZS metamaterial can be tailored with specific load-bearing capacities, making it suitable for vibration isolation across a range of applications.

\subsection*{\textbf{Static characteristics}}
In this section, we study the quasi-zero stiffness behavior of the function unit under static loading. As mentioned in the previous section, the QZS of the proposed shell is mainly influenced by its geometric parameters $h$, $t$, and $L$. Thus, varying any of them could lead to different stiffness responses. Here, we investigate the QZS property and explore the associated design space based on these geometric parameters.


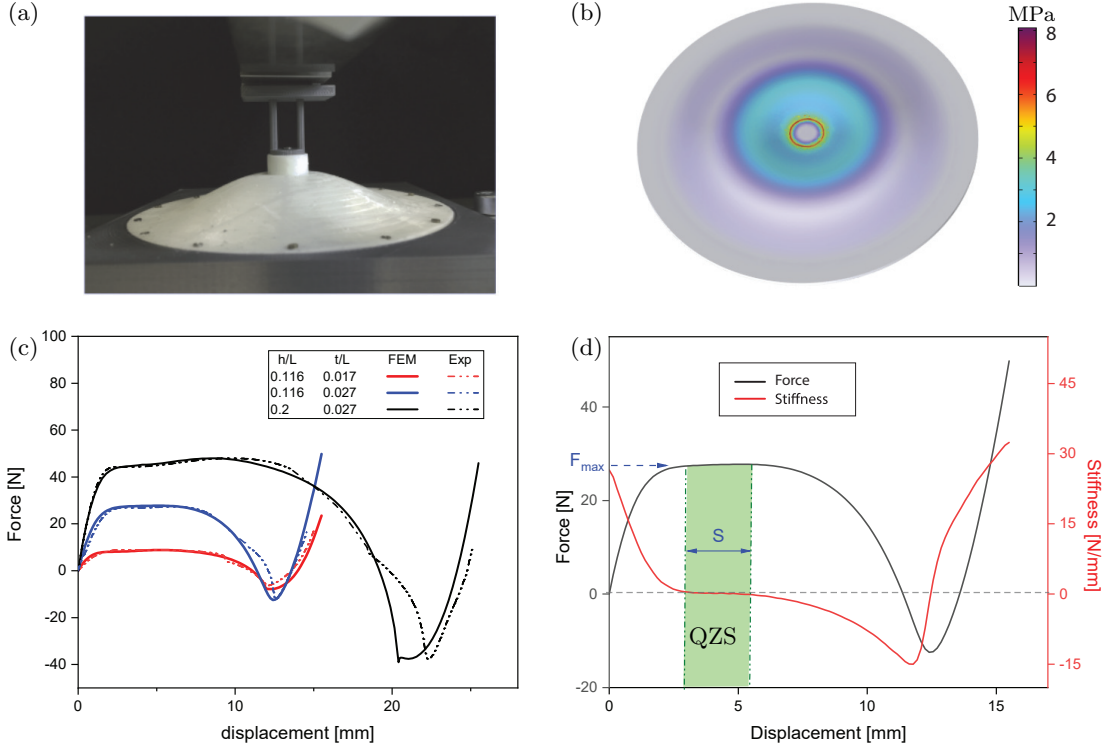
\begin{figure}[!ht]
	\centering
	\begin{tikzpicture}
	\node[anchor=south west,inner sep=0pt] (image) at (0,0) {\includegraphics[trim={0cm 0cm 0cm 0cm},clip,width=0.9\textwidth]{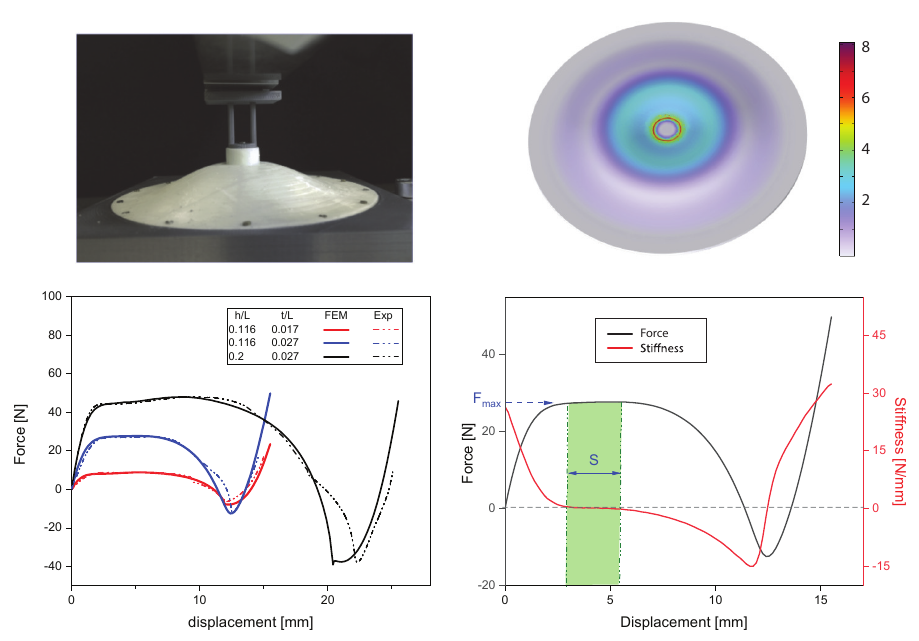}};
	\begin{scope}[x={(image.south east)},y={(image.north west)}]
	\node[align=center] at(0.03,0.95){\small (a)};
	\node[] at(0.53,0.95){\small (b)};
        \node[] at(0.925,0.95){\footnotesize MPa};
	\node[] at(0.03,0.52){\small (c)};
	\node[] at(0.53,0.52){\small (d)};
        \node[] at(0.642,0.15){\small QZS};
	\end{scope}
	\end{tikzpicture}
	\caption{Static response of the QZS unit. (a) The unit cell is loaded with a force applied in the center. (b) Von Mises stress of the shell during loading. (c) Measured and simulated force-displacement curves of multiple units with different geometric parameters. (d) Stiffness characteristics: $F_\mathrm{max}$ refers to the maximum load-bearing force; $S$ represents the width (i.e., displacement range) of the corresponding QZS region (in green).}
	\label{static}
\end{figure}
To characterize the static zero-stiffness behavior, we measure the shell's force-displacement curves ($F-d$) under uniaxial loading (see Fig.~\ref {static}(a)). The corresponding $F-d$ curves for different geometric parameters are presented in Fig.~\ref {static}(c). The results show that initially, the shell unit exhibits positive stiffness as displacement increases, which is associated with the shell's axisymmetric deformation as observed from both experiments and simulations (see Fig.~\ref {static}(b)). When the displacement reaches a certain value, the associated force increases to a maximum value, denoted as $F_\mathrm{max}$, at which the force remains approximately unchanged (i.e., the QZS region). The derived stiffness curve shown in Fig.~\ref {static}(d) further confirms the presence of the QZS region, where the stiffness is very close to zero. In this study, we take 0.1 N/mm as a threshold to define the QZS region, and the resulting displacement range of QZS region is denoted as $S$ (see Fig.~\ref {static}(d)). A larger $S$ is typically preferred since it facilitates maintaining low stiffness and resonance frequency in practical applications. Furthermore, as displacement continues to increase, the force eventually decreases to a negative value, indicating that this shell element can also exhibit bistable behavior \cite{qiu2004curved}. In this study, we mainly focus on the QZS region, which is essential for vibration isolation. {As the displacement increases further, the force-displacement curve exhibits a positive stiffness again and an increase in load. This load increase is mainly caused by the shell membrane's in-plane deformation.} Moreover, a good agreement between the experimental and numerical results is observed from Fig.~\ref {static}(c), for which discrepancies are mainly attributed to local deformations of the shell samples induced by geometric and clamping imperfections in practice.  

{Apart from maintaining a low stiffness, it is important for QZS elements to sustain a high $F_\mathrm{max}$, i.e., improving their load-bearing capability. In practical isolation applications, the QZS mechanism needs to withstand a certain payload. Here, we compare the load-bearing capability of the proposed QZS shells against beam-based designs that have been widely studied in the literature. To enable a reasonable comparison, a parameter $F_\mathrm{max}/(EV)$ is defined, where $E$ represents the material Young's modulus and $V$ represents the unit cell volume. Table 1 summarizes $F_\mathrm{max}/(EV)$ for different designs. As seen from the table, the proposed shell-based design outperforms the beam-based design in $F_\mathrm{max}$ with at least a factor of 2 improvement. }

\begin{table}\normalsize
    \centering
        \caption{\small{Comparison of load-bearing capability for different designs.}}
    \begin{tabular}{|p{5cm}|p{5cm}|}
    \hline
        \textbf{Design} & $F_{max}$/(EV) \enspace[$\times$ \num{e-5}]\\[20pt]
        \hline
         this work &  4.1\\
         \hline
         sinusoidal beam \cite{fan2020design,dalela2022design,dalela2024novel, liu2025straight,zhang2024design}& 1.0 -- 1.7\\
         \hline
         spline curved beam \cite{zhang2021tailored,hou2024quasi}& 0.5--0.6\\
         \hline
         arc-shaped beam \cite{zhou2024nonlinear}& 0.9 -- 1.2\\
         \hline
         inclined beam \cite{zhang2023bistable,cai2022flexural,cai2020design,liang2024design}& 0.8--1.3\\
         \hline
    \end{tabular}
    \label{tab:my_label}
\end{table}

\begin{figure}[!ht]
	\centering
	\begin{tikzpicture}
	\node[anchor=south west,inner sep=0pt] (image) at (-1,0) {\includegraphics[trim={0cm 0cm 0cm 0cm},clip,width=1\textwidth]{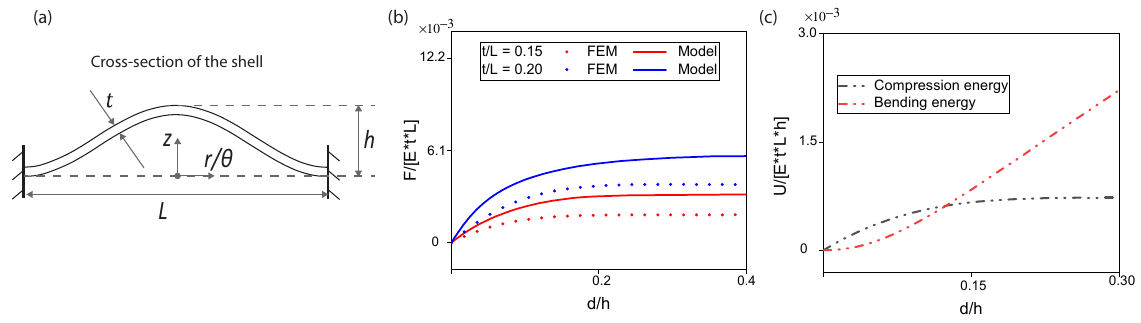}};
	\end{tikzpicture}
	\caption{The shell’s nonlinear static response from a simplified model. (a) Cross-sectional profile of the shell, where the shape is defined based on a polar coordinate system ($Z-\frac{r}{\theta}$). (b) Force-displacement (denoted as $F – d$) obtained from the model and simulations. (c) Compression and bending energy as a function of displacement.}
	\label{fig3-new}
\end{figure}

\subsection*{\textbf{Theoretical modeling}}

{To better understand the QZS mechanism, we build a simplified theoretical model based on the minimum potential energy principle. Figure~\ref {fig3-new}(a) illustrates the cross-section of the shell, where $Z$ -- $r$/$\theta$ represents a polar coordinate. Based on this, the shell’s axisymmetric undeformed shape ($Z_0$) can be expressed as:
\begin{equation}
{Z}_0 = \frac{H}{2} (1- \cos(2\pi \frac{r}{L})) \enspace \text{and}\enspace r \in[0,\frac{L}{2}], \enspace\theta\in[0,2\pi].\label{s1}
\end{equation}
Here, mode shapes $Z_1$, $Z_3$, and the shell’s deformed shape $Z$ are assumed as:}

{\begin{equation}
{Z}_i = 1 - \cos({N_i} \frac{r}{L}) \enspace \text{and} \enspace {N}_i = (i+1)\pi, \enspace \text{and} \enspace i= 1,3 \label{s2}
\end{equation}
\begin{equation}
{Z} = A_1 Z_1 + A_3 Z_3,\enspace \label{totao}
\end{equation}
where $A_1$ and $A_3$ are two unknown coefficients to be solved. Next, the shell's bending ($U_b$), compression($U_c$), and total strain energy ($U_s$) can be expressed as: 
\begin{equation}
\begin{cases}
U_b = \frac{Et^3}{24(1-v^2)}\int_{0}^{2\pi}\int_{0}^{L/2}(\frac{d^2Z}{dr^2} - \frac{d^2Z_0}{dr^2})^2 r drd\theta\\
\addlinespace
U_c = \frac{E\pi}{L}\int_{0}^{L/2}[(\frac{dZ}{dr})^2 - (\frac{dZ_0}{dr})^2]dr \cdot\int_{0}^{L/2}[(\frac{dZ}{dr})^2 - (\frac{dZ_0}{dr})^2 ]rdr\\
\addlinespace
U_s = U_b + U_c\label{eq}
\end{cases}
\end{equation}
The total potential energy ($U_t$) is thus expressed as a sum of $U_s$ and the potential energy ($U_p$) due to the load $F$, namely: \begin{equation} {U_p} = -F (H-2A_1).\enspace \label{totao} \end{equation}
Based on the minimum potential energy principle, a system of two equations can be derived and solved by taking the derivative of $U_t$ with respect to $A_1$ and $A_3$.}\\

{The analytical results are shown in Fig.~\ref {fig3-new}(b), and it can be seen that the non-linear deformation characteristics are well captured with this simplified theoretical model, leading to a positive stiffness followed by a ‘zero’ stiffness region (i.e., constant force). Included in the figure are the FEM results. The differences between the analytical and simulation results are mainly caused by the assumptions of small strain and no tangential compression adopted in the analytical modeling and the simplification of using only two mode shapes when solving the equations. In the future, it would be interesting to develop a full, accurate analytical model to explore this type of QZS shell design.\\}

{Using this analytical model, in Fig.~\ref {fig3-new}(c) we plot the bending and compression energy as a function of displacement. At the start of loading, the compression energy is dominant and increases rapidly, and the bending energy grows at a smaller rate. At a certain displacement (d/h = 0.12 in Fig. 3(c)), the bending energy becomes dominant in the energy landscape with a greater contribution than compression. More importantly, within a specific range (from 0.15 to 0.3), the bending energy shows approximately a linear relationship with displacement, while the compression energy remains almost constant. This indicates that the total strain energy increases linearly with displacement, leading to a constant force in the force-displacement curve, namely QZS behavior. Therefore, the QZS mechanism mainly arises from the bending energy, which dominates and increases linearly within this displacement range.}

\subsection*{\textbf{Parametric study}}
For vibration isolation, $F_\mathrm{max}$ and $S$ are two important parameters, because $F_\mathrm{max}$ determines the maximum payload one QZS unit can support and $S$ represents the effectiveness of realizing low-frequency isolation. To investigate the influence of geometric parameters on $F_\mathrm{max}$ and $S$, several samples with varying geometric parameters are tested. As shown in Fig.~\ref {static}(c), the force-displacement responses exhibit clear differences with varying geometric parameters. In the graph, $h$ and $t$ are normalized by $L$ to facilitate comparison. It can also be seen that increasing either $h/L$ or $t/L$ leads to a higher $F_\mathrm{max}$, while $S$ decreases as $t/L$ increases. To have a thorough understanding of the effects of $h$, $t$, and $L$, a parametric study is conducted in this section. 

\begin{figure}[!ht]
	\centering
	\begin{tikzpicture}
	\node[anchor=south west,inner sep=0pt] (image) at (-1,0) {\includegraphics[trim={0cm 0cm 0cm 0cm},clip,width=1\textwidth]{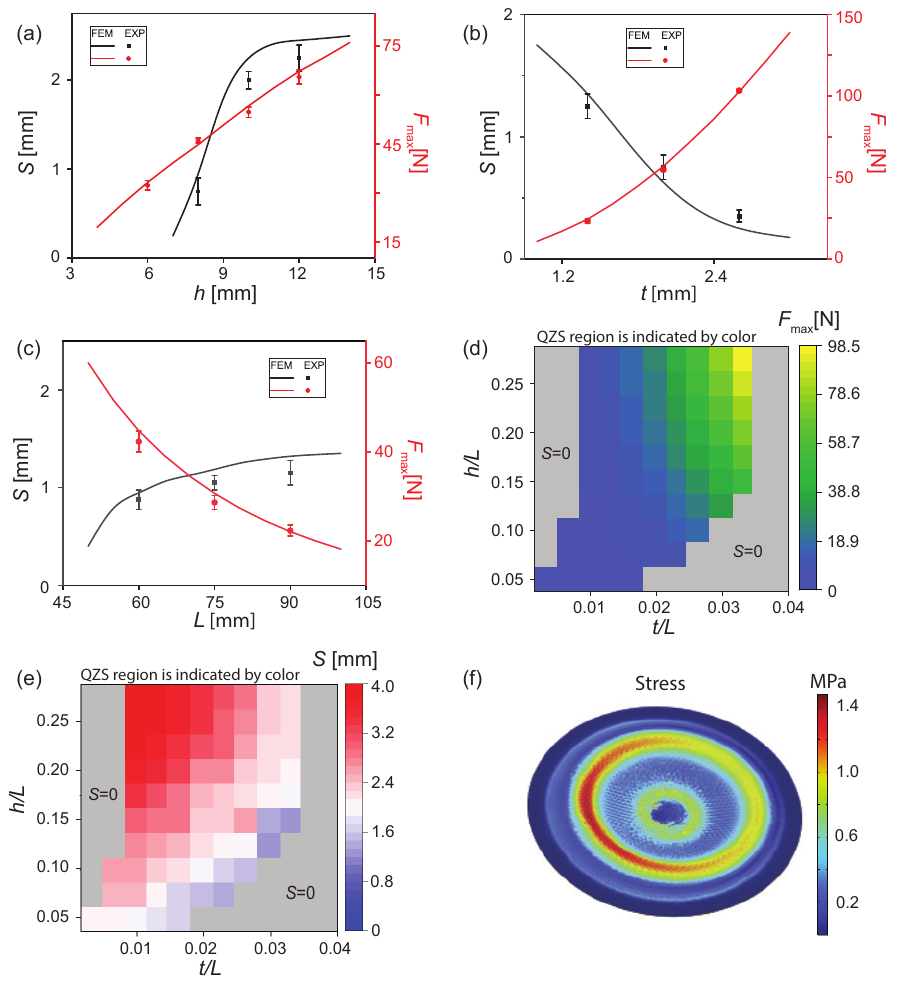}};
	\end{tikzpicture}
	\caption{The influence of geometric parameters on quasi-zero stiffness property. (a)--(c) Varying $h$, $t$, and $L$ result in different impact on $F_\mathrm{max}$ and $S$. {The black lines and dots represent the size of the QZS region, with the red color indicating the maximum force. The lines represent the FEM results, while the dots denote the experimental data.} (d)--(e) {Simulation results showing design space for QZS behavior: QZS is only effective within the colored region, while values of $h/L$ and $t/L$ in the grey region indicate that QZS cannot be realized. (f) Simulated stress distribution resulting from the deformation of a thin shell.}}
	\label{fig3}
\end{figure}

The testing results of $S$ and $F_\mathrm{max}$ as a function of $h$ while fixing $t$ and $L$ are shown in Fig.~\ref {fig3}(a). Both experiments and simulations indicate that increasing $h$ leads to a linear increase of $F_\mathrm{max}$. This can be explained by the fact that $h$ determines the initial curvature of the shell, and a larger $h$ causes a higher bending and compression energy during loading, thereby leading to a higher force output. Moreover, a nonlinear relation between $S$ and $h$ is observed in this figure. As $h$ increases, $S$ quickly increases but eventually converges to a maximum value of about 2.3 mm in this case. Beyond this point, a further increase in $h$ does not significantly enhance $S$ due to the shell's nonlinear characteristics. Notably, when $h$ is reduced, $S$ approaches zero, indicating that QZS behavior is not valid ($S=0$) for a small $h$. Therefore, to design shells that exhibit QZS, $h$ must exceed a certain threshold value. 

A similar study is undertaken for the thickness $t$. As depicted in Fig.~\ref {fig3}(b), $F_\mathrm{max}$ increases dramatically with changes in thickness, with a steeper increase compared to $h$. {As shown in Equation (5), the bending stiffness is expressed as $\frac{Et^3}{12(1-v^2)}$, and $h$ contributes to the bending energy in the form of $h^2$.} Therefore, tuning $t$ is more effective than $h$ for cranking up the shell's load-bearing capability. Regarding $S$, a smaller $t$ results in a wider QZS region, while increasing $t$ causes $S$ to decrease toward zero, indicating the loss of QZS property. Thus, to enable QZS behavior, $t$ should be constrained to be smaller than a critical value. In addition, Fig.~\ref{fig3}(c) displays the change of $F_\mathrm{max}$ and $S$ as a function of the length $L$. It shows that $F_\mathrm{max}$ decreases with the increase of $L$, because a larger $L$ results in a more compliant shell. Compared to $t$, the impact of $L$ on both $F_\mathrm{max}$ and $S$ is relatively minor.

The above results show that varying geometric parameters not only affects $F_\mathrm{max}$ but also determines whether the QZS behavior can be realized. To properly design a QZS unit cell, $h, t, L$ must be specified within an appropriate range. Using the verified finite element models, we explore the design space that allows for QZS properties. Specifically, the maximum force $F_\mathrm{max}$ and QZS region $S$ are calculated by varying the normalized parameters $h/L$ and $t/L$. The simulated results are presented in Fig.~\ref {fig3}(d)--(e), and two distinct regions are identified: i) Grey region represents that the shell does not exhibit QZS behavior ($S=0$); ii) Colored region corresponds to the design space where QZS is achievable (i.e., $S>0$). For instance, when $h/L = 0.15$, $t/L$ should be within the range of [0.01-0.033] to realize QZS. If $t/L$ is either below 0.01 or above 0.033, the shell fails to achieve a zero-stiffness plateau in the force-displacement curve. Figure~\ref {fig3}(f) displays the stress distribution of a shell with $t/L<0.01$. It is evident that the shell's deformation is not axisymmetric, and in this example, stress on the left side is higher than on the right side. This non-axisymmetric deformation occurs because a shell with a small $t$ is prone to local buckling, which prevents it from exhibiting zero-stiffness properties. Moreover, to enlarge the QZS range ($S$), a high $h/L$ ratio is preferred, as shown in Fig.~\ref {fig3}(e). For cranking up $F_\mathrm{max}$, both $h$ and $t$ need to be increased. Furthermore, the results in Fig.~\ref {fig3}(d) confirm that $t/L$ has a more significant impact on $F_\mathrm{max}$ compared to $h/L$, causing $F_\mathrm{max}$ to vary over a large range. Therefore, it is crucial to tweak $t/L$ to enhance the load-bearing capability.

\subsection*{\textbf{Dynamics characteristics}}

The QZS property discussed above demonstrates that the proposed QZS functional unit can realize a very low resonance frequency when its deformation reaches the QZS region. In terms of vibration isolation performance, a low resonance frequency results in a low transmittance, meaning that external vibration disturbance will be minimally transmitted to the object to be isolated. To evaluate the dynamic performance of the metamaterial, frequency and transmission tests are undertaken with various payloads.

\begin{figure}[!ht]
	\centering
	\begin{tikzpicture}
	\node[anchor=south west,inner sep=0pt] (image) at (-2,0) {\includegraphics[trim={0cm 0cm 0cm 0cm},clip,width=1\textwidth]{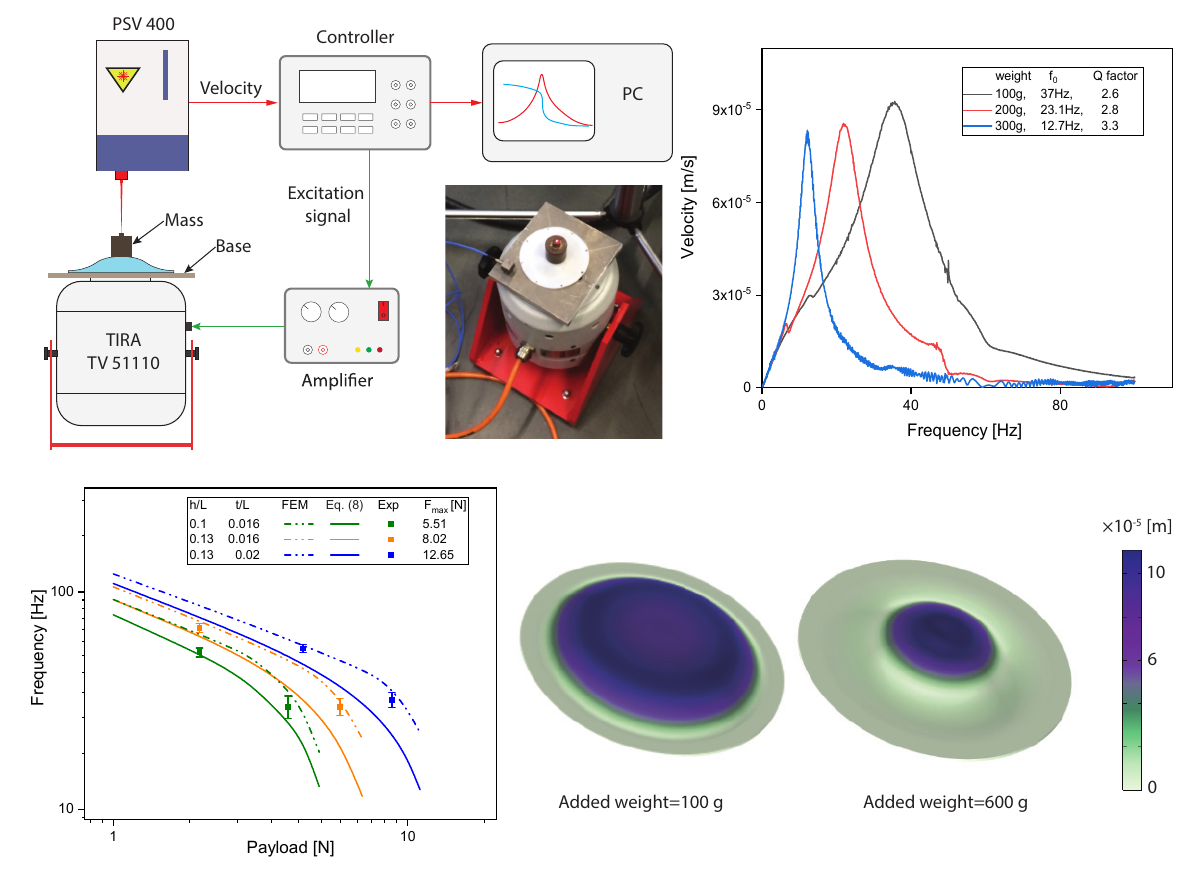}};
	\begin{scope}[x={(image.south east)},y={(image.north west)}]
	\node[] at(0.03,0.95){\small (a) };
	\node[] at(0.36,0.8){\small (b) };
	\node[] at(0.7,0.95){\small (c) };
	\node[] at(-0.04,0.45){\small (d) };
	\node[] at(0.43,0.45){\small (e) };
        \node[] at(0.68,0.45){ mode shape };
	\end{scope}
	\end{tikzpicture}
	\caption{Dynamical characterization of the proposed QZS structures. (a) Schematic of the experimental setup for dynamics measurement. The laser is pointed at the mass and base to measure their vibration, respectively. (b) A picture showing a mass supported by the QZS unit cell in the test. (c) {Measured frequency response curves for three situations: added weight of 100 g, 200 g, and 300 g. The legend also provides the extracted resonance frequencies and the corresponding  Q factors for each peak. (d) Measured (Exp) and simulated (FEM) first resonance frequency as a function of payload (i.e., added mass) for shells with different geometry parameters. Applied payloads are lower than $F_\mathrm{max}$ to avoid dynamic instability. Results derived from Equation (8) are also displayed (see solid lines).  (e) Simulated mode shapes of a shell when a mass of 100 g and 600 g are added, respectively.}}
	\label{Model}
\end{figure}
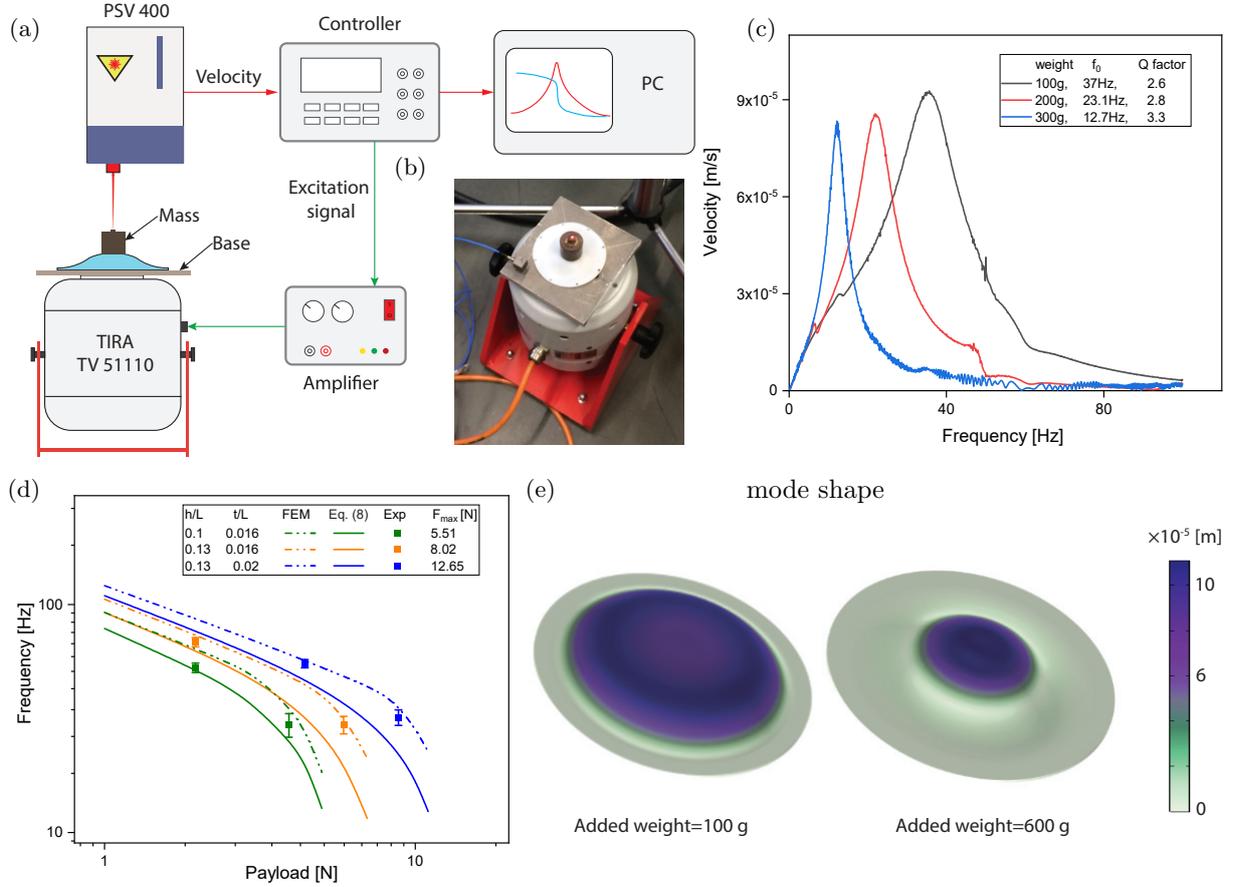

The dynamic response of the shells is measured using the experimental setup as shown in Fig.~\ref{Model}(a). As shown in this figure, the shell sample is mounted onto a shaker, which is connected to a power amplifier. The excitation signal is generated by a controller linked to a PC. To detect the shell's dynamic response, we use a Polytec Laser Doppler Vibrometer (LDV) to measure the out-of-plane velocity of the object. The LDV laser beam is focused on points of interest, and the vibrometer signal obtained by the laser head is analyzed using a Polytec decoder. Additionally, to test the transmittance performance, a mass is put on the top of the metamaterial to compress the shell close to its QZS region (see Fig.~\ref {Model}(b)).

{The measured frequency resonance curves of a QZS unit loaded with different weights are shown in Fig.~\ref {Model}(c). In this study, only the first resonance frequency ($f_0$) is studied since only the low-frequency response is of interest for QZS vibration isolators. In Fig.~\ref {Model}(c), a clear resonance peak is observed for each payload and those resonance peaks can be described by a single degree-of-freedom lumped oscillator:
\begin{equation}
    m\Ddot{x}+c\Dot{x}+k_\mathrm{static}x=F(t),
    \label{eq:resonator}
\end{equation}
where, $x$ is the displacement; $m,c,k_\mathrm{static}$ are the effective mass, damping coefficient, and static stiffness of the shell, respectively; $F(t)$ is the applied force. By solving Eq. (\ref{eq:resonator}) and fitting it with the frequency response curves, their resonance frequencies can be obtained and shown in the legend of Fig.~\ref {Model}(c). From Fig.~\ref {Model}(c), we find that when the added mass increases from 100 g to 300 g, the resulting resonance frequency decreases from 37 Hz to 12.7 Hz. This frequency reduction is expected since the resonance frequency is expressed by: 
\begin{equation}
    f=\frac{1}{2\pi}\sqrt{\frac{k_\mathrm{static}}{m}}.
\end{equation}
When the structure's effective mass $m$ increases with the payload, the resonance frequency drops. Also, as presented in the previous section (Fig.~\ref{static}(d)), the shell's stiffness gradually decreases with increased force before reaching $F_\mathrm{max}$ ($F_\mathrm{max} = 4.8$ N in this case). Thus, it is expected that adding a heavier mass, approaching approximately 480 g, will cause the corresponding resonance frequency to decrease towards zero. }

To quantify the relationship between the resonance frequency and added mass, a few samples with different geometric parameters are tested both experimentally and numerically.  {The results are shown in Fig.~\ref {Model}(d), where the resonance frequency is plotted as a function of the added mass on a log-log scale. It is noted that for small payloads (less than 2 N), the frequency-payload curves show a slope of -0.55, which is slightly steeper than the theoretical value of -0.5 predicted by Eq. (8). This is because the stiffness is not a constant value with respect to the payload; instead it decreases during loading, as shown in Fig.~\ref{static}(d). In addition, when a payload is higher than a certain value, a deep slope in the frequency-payload curve is observed, where the stiffness reduction plays an important role. As demonstrated in Fig.~\ref{static}(d), static stiffness ($k_\mathrm{static}$) quickly decreases as a function of payload, resulting in a rapid decrease of resonance frequency.} In addition, Fig.~\ref {Model}(d) shows that the frequency-mass relation is highly dependent on geometric parameters. {Specifically, smaller values of $h/L$ and $t/L$ lead to greater sensitivity of the frequency to variations in mass. To better understand the influence of the added mass on the shell's dynamics, we also simulate its mode shapes.} Figure~\ref{Model}(e) plots the simulated mode shapes of a shell with an added mass of 100 g and 600 g, respectively. These mode shapes indicate that a heavier mass (i.e., higher payload) makes the shell's vibration concentrate locally on the central region. This is because when the added mass is small, the shell's static deformation is minimal, allowing for a large vibration region. However, when the added weight is bigger, the shell undergoes significant pre-deformation, leading to stress concentration in the central region, thus making it more compliant (see Supplementary Figure 1).

{
Apart from resonance frequencies, damping is another important characteristic of the shell's dynamics. By fitting the curves in Fig.~\ref{Model}(c), we also obtain their corresponding Q factors, which quantify the energy dissipation of each resonance mode. These values are provided in the figure legend. We can see that the Q factors are only 2-3 under different loads. The measured Q factors reflect the overall damping originating from all energy dissipation sources. We attribute the primary energy dissipation to material damping, as the material used in our study is thermoplastic elastomers (TPU), which are found to have significant viscoelastic damping  \cite{medel2022stiffness,zhang2021novel,liu2024compact,zu2025tailor}. When the TPU shell vibrates, energy is dissipated as heat due to the hysteresis inherent in its viscoelastic property. Another dominant source is likely the air damping due to the squeezed film effect, which is a common damping source when a structure is vibrating close to another substrate in air \cite{chen2024nonlinear,bao2007squeeze}. In our case, air is confined within the shell, making the air damping more significant. Additionally, friction in the clamping structure may also contribute to the damping in the shell structure (see Fig.~\ref{Model}(b)). To better understand the damping mechanisms, future work is required to develop a more sophisticated model that can quantitatively characterize the various sources of energy dissipation.
}

\subsection*{\textbf{Transmittance and isolation performance}}

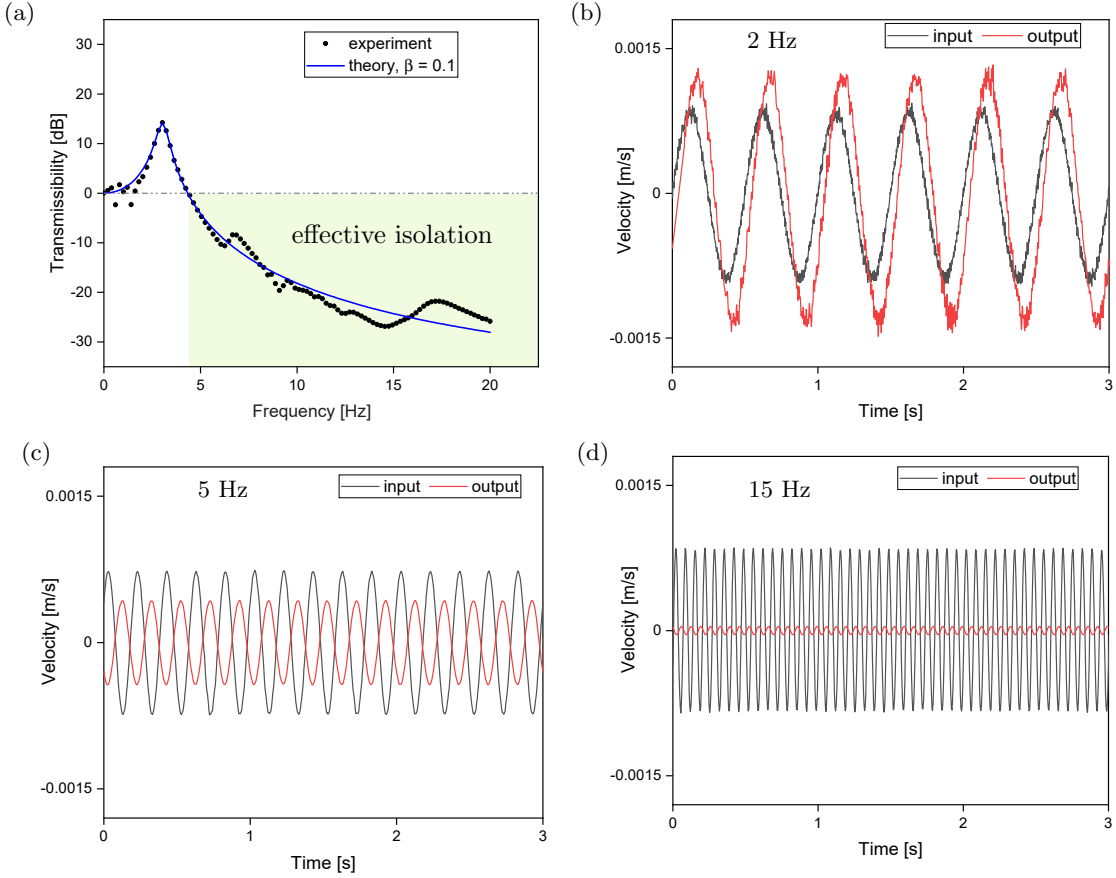
\begin{figure}[!ht]
	\centering
	\begin{tikzpicture}
	\node[anchor=south west,inner sep=0pt] (image) at (-1.5,0) {\includegraphics[trim={0.2cm 0cm 0cm 0cm},clip,width=0.9\textwidth]{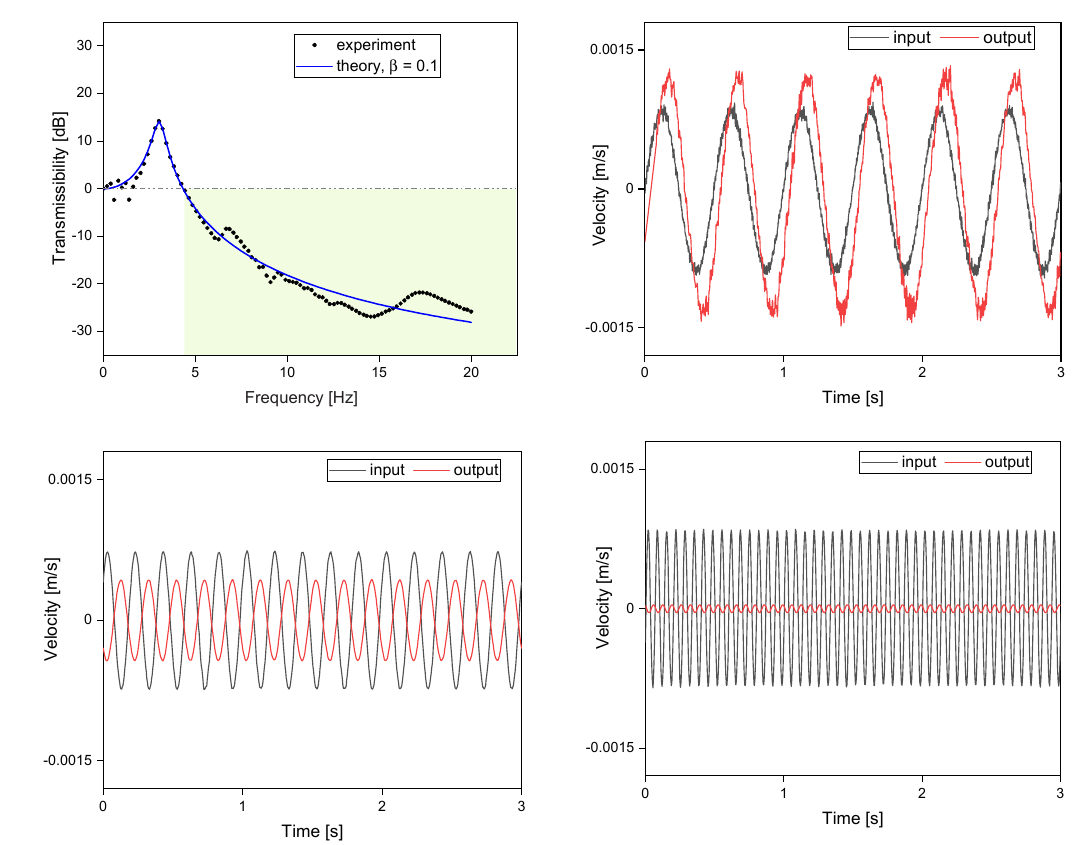}};
	\begin{scope}[x={(image.south east)},y={(image.north west)}]
	\node[] at(0.01,0.98){\small (a) };
        \node[] at(0.35,0.73){effective isolation};
	\node[] at(0.57,0.98){\small (b) };
    \node[] at(0.75,0.95){\small 2 Hz };
    \node[] at(0.15,0.44){\small 5 Hz };
    \node[] at(0.7,0.44){\small 15 Hz };
	\node[] at(-0.03,0.48){\small (c) };
      \node[] at(0.52,0.48){\small (d) };
	\end{scope}
	\end{tikzpicture}
	\caption{Vibration isolation performance of the QZS unit. (a) Measured and theoretical transmittance curve for a unit cell with an added mass of 280 g. (b)--(d) Measured dynamic response of the base (input) and mass (output) for different frequencies (2 Hz, 5 Hz, and 15 Hz). }
	\label{fig5}
\end{figure}

To verify the vibration isolation performance, we measure the transmissibility using the experimental setup illustrated in Fig.~\ref{Model}(a). In particular, a QZS unit with a known static $F_\mathrm{max}$ is compressed by a dedicated mass whose gravity force is close to $F_\mathrm{max}$. We then perform a frequency sweep over the range of 0-20 Hz, during which the base velocity of the shaker (input) and the velocity of mass on top (output) are recorded by LDV laser beam, respectively (see Fig.~\ref {Model}(a)). The transmissibility ($T$), expressed in decibels (dB), is defined as: 

\begin{equation}
{T} =20\times\log_{10}(A_\mathrm{out}/A_\mathrm{in}), \label{shape_initial}
\end{equation}
where $A_\mathrm{out}$, $A_\mathrm{in}$ represent the measured output and input velocity, respectively. The measured transmittance is presented in Fig.~\ref {fig5}(a), where the tested QZS unit exhibits a static $F_\mathrm{max}$ of 3.1 N, and a selected mass of 280 g is added as a payload. A resulting resonance frequency of 3 Hz is detected, and it can be noted that when the vibration frequency exceeds 3 Hz, transmissibility $T$ starts to quickly decrease into a negative value, meaning that the output amplitude is smaller than the input (i.e., $A_\mathrm{out} < A_\mathrm{in}$). In this case, the frequency corresponding to 0 dB is around 4 Hz, indicating that disturbances with frequencies larger than 4 Hz can be effectively isolated by the shell unit. As frequency increases further, $T$ decreases and reaches as low as $\SI{-20}{dB}$, representing that the input vibrations are largely attenuated. For a linear spring isolation, its transmissibility as a function of frequency ($w$) can be described as:

\begin{equation}
{T} =20\times \log_{10}(\sqrt{\frac{1+(2\beta w/w_0)^2}{(1-(w/w_0)^2)^2+(2\beta w/w_0)^2}}) \quad \text{and} \quad \beta = \frac{c}{2\sqrt{K/M}},  \label{new}
\end{equation}
where $w_0$ and $\beta$ represent the system's first eigenfrequency and damping ratio, respectively. $K$ and $M$ represent the effective stiffness and mass. For the shell unit, this analytical relation with $\beta$ of 0.1 is displayed in Fig.~\ref {fig5}(a), demonstrating a good agreement with experimental data. Moreover, Eq.~(\ref{new}) suggests that increasing $h/L$ or $t/L$ would improve the isolation performance by lowering the frequency at which the transmissibility reaches zero-dB. This is because from the parametric study in the previous section, it has been noted that enlarging $h/L$ or $t/L$ can result in a higher $F_\mathrm{max}$, thus a bigger mass $M$ for vibration isolation. As a result, the associated eigenfrequency $w_0$ will decrease, and the peak in Fig.~\ref {fig5}(a) would further shift to the left, leading to improved vibration isolation performance. 

The isolation performance is also demonstrated in Fig.~\ref {fig5}(b)--(d), which show the measured velocity responses of the input and output in the time domain at 2 Hz, 5 Hz, and 15 Hz, respectively. It can be seen that for the case of 2 Hz, the input velocity is amplified so that a higher output velocity is observed. This amplification is most evident near the resonance frequency around 3 Hz, as shown in Fig.~\ref {fig5}(a). On the other hand, time domain responses for cases of 5 Hz and 15 Hz demonstrate the shell's isolation capability, showing a very low amplitude vibration of the isolated object. {In addition, it should be noted that TPU plastics normally have a specific lifetime and viscosity \cite{neubauer2022dma}, which may lead to property and performance changes over time. In future work, to strengthen the fidelity of the proposed shell elements, their QZS isolation behavior should also be tested with multiple loading cycles to evaluate the stability and repeatability.}


\begin{figure}[!ht]
	\centering
	\begin{tikzpicture}
	\node[anchor=south west,inner sep=0pt] (image) at (-1.5,0) {\includegraphics[trim={0cm 0cm 0cm 0cm},clip,width=1\textwidth]{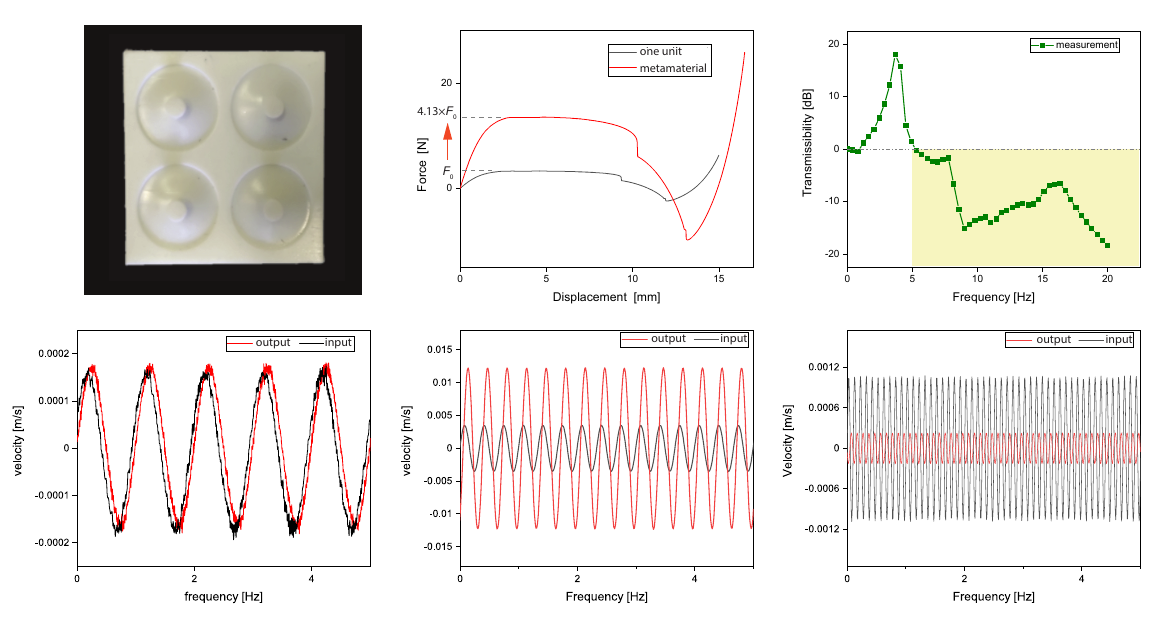}};
	\begin{scope}[x={(image.south east)},y={(image.north west)}]
	\node[] at(0.04,0.95){\small (a) };
	\node[] at(0.4,0.95){\small (b)};
        \node[] at(0.96,0.73){\footnotesize effective isolation};
	\node[] at(0.75,0.95){\small (c)};
        \node[] at(0.1,0.44){\small 1 Hz };
        \node[] at(0.45,0.44){\small 3 Hz };
        \node[] at(0.8,0.44){\small 10 Hz };
	\node[] at(0,0.48){\small (d)};
        \node[] at(0.35,0.48){\small (e)};
        \node[] at(0.7,0.48){\small (f)};
	\end{scope}
	\end{tikzpicture}
	\caption{Vibration isolation performance of a QZS metamaterial. (a) A metamaterial sample consisting of four QZS units (2$\times$2 units). (b) Measured static force-displacement curves of a QZS unit (maximum load denoted as $F_0$) and the metamaterial with 2$\times$2 units. (c) Experimentally measured transmittance curve of the metamaterial, for which a mass of 1100g is added. (d)--(f) Measured dynamic behavior: vibration velocities of the base (input) and mass (output) are detected at frequencies 1 Hz, 3 Hz, and 10 Hz, respectively.}
	\label{space}
\end{figure}
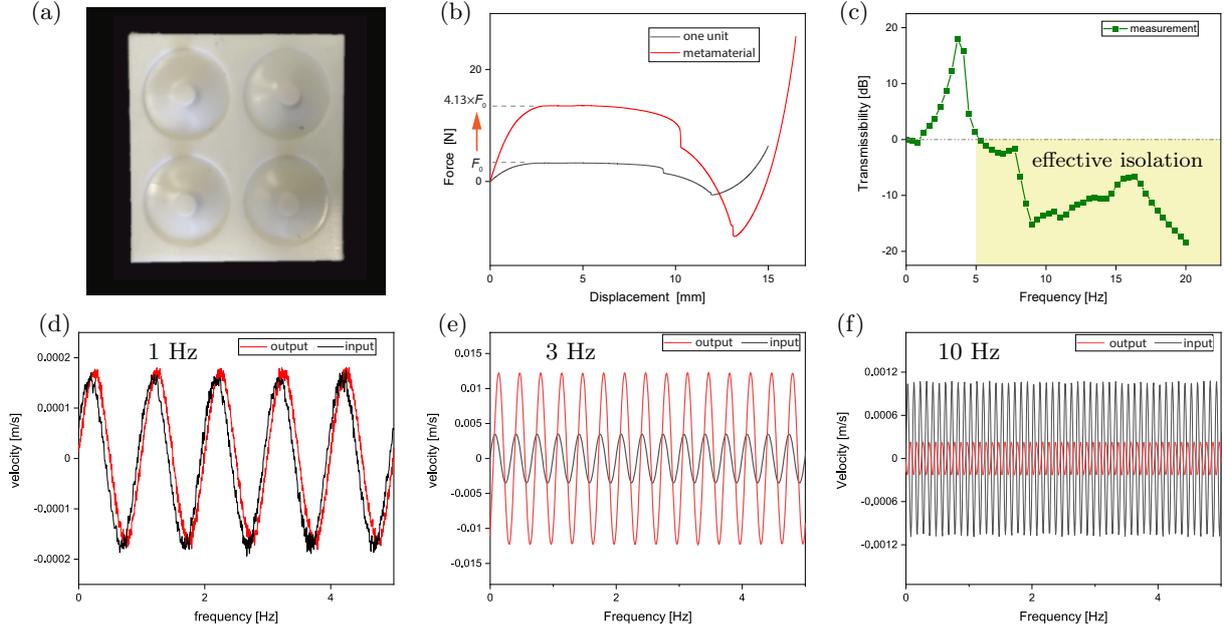


The zero stiffness and associated low-frequency resonance of the QZS unit arise from their structural nonlinear deformation, eliminating the need for assembling positive and negative stiffness parts. This inherent design simplicity significantly reduces the complexity of vibration isolation systems. Consequently, these functional units can be arranged in various patterns as metamaterials to accommodate different payloads. Here, we demonstrate and measure the vibration isolation of a printed metamaterial consisting of four QZS units (2 $\times$ 2 arrangement in X-Y direction), as shown in Fig~\ref {space}(a). The static compression result of this metamaterial is presented in Fig.~\ref {space}(b), confirming that an explicit QZS region is well maintained. Notably, the measured $F_\mathrm{max}$ of the metamaterial is about 4.13 times higher than that of a single unit, which is very close to the theoretical value of 4. The small deviation is primarily due to the geometric imperfections in the printed shell samples. Moreover, similar vibration tests are conducted for this metamaterial, with an added payload of 1100 g placed on top. The measured transmissibility results are presented in Fig.~\ref {space}(c). {
From Fig.~\ref {space}(c), a dominant resonance peak is observed around 4 Hz. Additionally, smaller peaks appear around 7 Hz and 16 Hz, likely due to uneven forces acting on the four shells as a result of imperfect loading conditions. This change in the boundary conditions leads to a complex frequency response. Besides, the sudden amplitude drop around 7 Hz may be attributed to the insufficient frequency resolution and another more dominant peak around 16 Hz, creating a frequency valley. In terms of vibration isolation, it can be seen that this metamaterial exhibits good vibration isolation performance while maintaining higher load-bearing capability. Specifically, when the actuation frequency is higher than 5 Hz, the transmissibility turns into a negative value ($A_\mathrm{out} < A_\mathrm{in}$), indicating an effective vibration isolation.} In addition, the dynamic responses measured at 1, 3, and 10 Hz also confirm the four-element structure's isolation capability in a low-frequency range (see Fig.~\ref {space}(d)--(f)). 

\begin{figure}[!ht]
	\centering
	\begin{tikzpicture}
	\node[anchor=south west,inner sep=0pt] (image) at (0,0) {\includegraphics[trim={0cm 0cm 0cm 0cm},clip,width=0.95\textwidth]{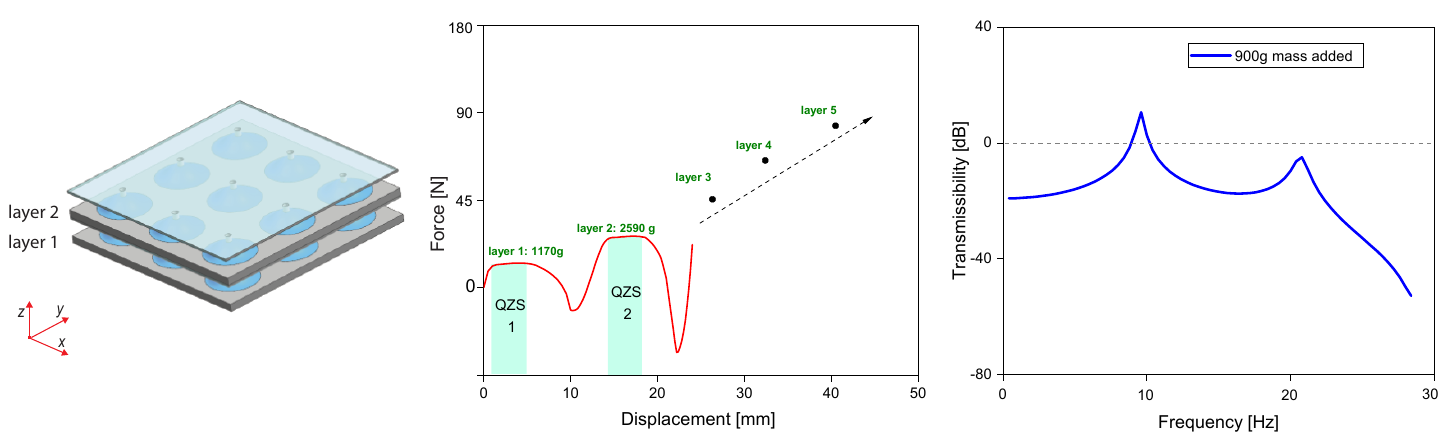}};
	\begin{scope}[x={(image.south east)},y={(image.north west)}]
	\node[] at(0.05,0.95){ \small (a) };
	\node[] at(0.28,0.95){\small (b) };
        \node[] at(0.66,0.95){\small (c) };
	\end{scope}
	\end{tikzpicture}
	\caption{Demonstration of a 3D QZS metamaterial for vibration isolation. (a) Illustration of a two-layer structure composed of 3 $\times$ 3 elements at each layer, where the shell thickness $t$ in the second layer is bigger than that of the first layer. The transparent plate on the top is used for putting the object to be isolated. (b) Static force-displacement curve of the two-layer metamaterial: two distinct zero-stiffness regions (QZS 1 and 2 highlighted in green) are observed, indicating it can support a weight of 1170 g and 2590 g, respectively. It can be expected that adding more layers (e.g., layer 3, 4, 5) with controlled shell geometric parameters will enable additional QZS regions with increased $F_\mathrm{max}$. {(c) Simulated transmission curve of the multi-layer metamaterial, where a mass of 900g is added as payload.}}
	\label{demo-1}
\end{figure}
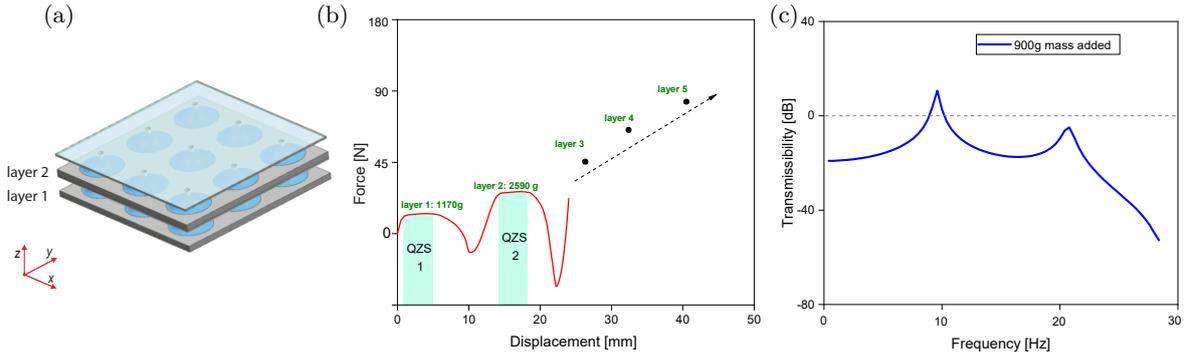

In addition to planar arrangements, the QZS units can also be combined in series, i.e., connecting multiple units vertically.  Figure~\ref {demo-1}(a) illustrates a 3D metamaterial that integrates both parallel (X-Y) and serial (Z) arrangements of the shell units. In particular, it consists of 3 $\times$ 3 unit cells in X-Y plane and two layers in Z, thus 18 elements in total. Note that different from planar arrangements, the serial arrangement allows for combining multiple layers with different static characteristics. For example, the QZS units in Layer 1 and 2 could be designed with different geometric parameters, enabling different $F_\mathrm{max}$ for each layer. In this example, nine shell elements in Layer 2 are designed with a higher thickness $t$ than those shells in Layer 1. The simulated static behavior of this metamaterial is displayed in Fig.~\ref {demo-1}(b), demonstrating that it is capable of exhibiting two sequential QZS regions. In particular, during the vertical compression, nine shells in Layer 1 first deform with a zero-stiffness response (denoted as QZS 1) since these shells have a smaller thickness and thus are more compliant than Layer 2. As the displacement increases further, the shell units in Layer 1 deform into a concave shape, and the nine shells of Layer 2 start to deform, leading to a second zero-stiffness plateau (denoted as QZS region 2) with a higher $F_\mathrm{max}$. The two sequential QZS regions here indicate that this two-layer metamaterial can isolate vibrations for two different payloads corresponding to 11.7 N and 25.9 N, respectively. Therefore, the two-layer mechanism cannot only isolate external vibrations for a mass of \SI{1170}{g}, but also a mass of \SI{2590}{g}. Furthermore, additional layers, e.g., Layer 3, 4, and 5, can be incorporated to form a five-layer metamaterial. As a result, five QZS regions can be expected in the static force-displacement curve (see Fig.~\ref {demo-1}(b)). By carefully selecting geometric parameters for the shells in each layer, it is possible to create more QZS regions, thereby accommodating a range of payloads and enhancing vibration isolation performance. {To verify the isolation performance of the multi-layer matematerial, we build a FEM dynamics model to calculate its transmissibility. Specifically, a multi-layer geometry is constructed, and the associated transmissibility response related to 900 g payload is shown in Fig.~\ref{demo-1}(c). Unlike a single-layer shell with only a dominant resonance peak, this multi-layer system exhibits two significant peaks at 9 Hz and 20 Hz, respectively. The first peak corresponds to the first layer of the metamaterial, which is compressed close to its QZS region. The second peak is attributed to oscillations in other layers, because stacking multiple layers in vertical directions results in a multi-degree-of-freedom system. Nevertheless, the second peak remains below 0 dB, indicating that effective isolation is achieved and demonstrating the isolation potential of such multi-layer systems.}

As demonstrated, arranging the shells in X-Y-Z directions paves the way for designing a multi-layer 3D metamaterial isolator capable of supporting multiple payloads and achieving programmable vibration isolation characteristics. In particular, based on the design space defined in Fig.~\ref {fig3}(d)--(e) and given the payload requirement, one can easily select appropriate geometric parameters for shell units in each layer. Then, a 3D QZS metamaterial composed of multiple shells in X-Y-Z directions can be generated to function as a vibration isolator to accommodate different payloads. This programmable strategy can integrate various vibration isolation characteristics into a single monolithic structure, thereby greatly reducing the need to design multiple mono-parts. {Moreover, the proposed shell geometry can be further extended to design acoustic metamaterials, for which the bandgap is an important performance indicator (see Supplementary Figure 6 for more details). A future study will be conducted to study the application of the proposed shells as acoustic metamaterials.}  

\section*{\textbf{Discussion}}

{Besides the quasi-zero-stiffness property, the proposed shell also exhibits negative stiffness at a certain displacement range (see Fig.~\ref{static}). It has been reported in literature that this negative stiffness is beneficial for energy absorption and motion-related applications \cite{bertoldi2017flexible, shan2015multistable}. Regarding vibration isolation, this negative stiffness might be a performance-limiting factor. }

{Depending on the payload level as well as the excitation amplitude, when the relevant dynamic load is smaller than $F_\mathrm{max}$, the shell can function well as a low-frequency isolator. However, in a static force-controlled loading, when the applied force is larger than $F_\mathrm{max}$, the negative stiffness would result in a snap-through displacement jump to another stable state. From the perspective of vibration isolation, the negative stiffness region indicates that the system would experience dynamic instability and jump to a new position, which in turn undermines isolation performance. This performance degradation is mainly caused by the high stiffness of the new stable position, leading to a high resonance frequency and making it not suitable for low-frequency isolation (see Supplementary Figure 3 -- 5). As a result, QZS region can no longer be maintained.  To ensure reliable operation, the excitation amplitude should be limited to guarantee the total dynamic load not exceeding $F_\mathrm{max}$. In addition, the negative stiffness region can be avoided by changing the design, e.g., adding an end stop to constrain the vibration range. Future study needs to be conducted to minimize the impact of negative stiffness.} 

In summary, we propose and investigate a type of shell-based metamaterial exhibiting quasi-zero stiffness characteristics. Instead of relying on the combination of a negative stiffness element with a positive stiffness counterpart, the proposed shell unit realizes a zero-stiffness property due to its inherent geometric nonlinearity, leading to a simple mechanism design for vibration isolation. Our findings reveal that QZS is highly dependent on the shell's geometric parameters, i.e., height ($h$), thickness ($t$), and length ($L$). Experimental and numerical results show that for realizing QZS, $t/L$ and $h/L$ have to be tuned within a certain range, and a very high $t/L$ or low $h/L$ will disable the QZS characteristics. Moreover, we study the dynamics of the proposed QZS metamaterial and demonstrate that by adding an appropriate payload, the metamaterial can achieve a very low resonance frequency. This characteristic makes it highly effective for isolating low-frequency vibration disturbances, thereby minimizing vibrations transmitted to the object. Additionally, we show that the proposed shell design could act as a building block for constructing vibration isolation metamaterials to accommodate multiple payloads. By arranging multiple shell elements in parallel and series, the 3D metamaterial can achieve several QZS regions associated with different payloads. This approach allows for the integration of diverse vibration isolation characteristics into a single monolithic structure. With a proper arrangement and design of such shell elements, the proposed concept can be further explored to develop functional isolators with tunable properties.

\section*{Methods}
\subsection*{\textbf{Fabrication and experiments}}
In this work, shell-type samples are printed with thermoplastic elastomers (TPU) using a fused deposition printer (Prusa i3 Mk2), and then they are connected with stiff frames made from polylactide material. To determine the TPU material properties, standard tensile tests are conducted on dumbbell specimens fabricated based on the ASTM D638-14 standard. The measured Young's modulus and Poisson's ratio are 95 MPa and 0.4, respectively.

Uniaxial loading tests are carried out to investigate the static behavior of the proposed design. A mechanical testing system (Zwick-005) is used to exert vertical loads in a displacement-controlled manner with a speed of 10 mm/min. Rigid connectors are adopted to connect the tensile machine head with the specimens, as shown in Fig.~\ref {static}(a). Moreover, to test the dynamic behavior of the proposed structure, a TIRA TV 51110 shaker is used to excite the structure into vibration with a chirp signal generated by the Polytec controller and a power amplifier. The resulting motion is read out by a Polytec PSV 400 laser Doppler vibrometer and subsequently transferred to a PC for vibration analysis. More details about the experimental setup for the dynamics test can be found in Supplementary Figure 2. 

\subsection*{\textbf{Numerical simulations}}
Nonlinear finite element models (FEM) are built in COMSOL to investigate the static and dynamic behavior of the proposed structure. Specifically, measured Young's modulus and Poisson's ratio are imported into FEM with a linear elastic constitutive relation, and a mesh convergence study is conducted to ensure accuracy. For static simulations, loads are applied at the center flat region ($w_1$), while the boundary region ($w_2$) is fully constrained. For dynamics simulations, a two-step analysis is performed, where the shell structure is first loaded with a static force, and then an eigenfrequency analysis is conducted to extract the corresponding resonance frequency and mode shape. 


\section*{Data availability}
The datasets used and/or analyzed during the current study are available from the corresponding author upon reasonable request.

\section*{Acknowledgements}
This research/project is supported by the National Research Foundation, Singapore, under its Competitive Research Programme (CRP Award No: NRF-CRP30-2023-0002).
Any opinions, findings and conclusions or recommendations expressed in this material are those of the author(s) and do not reflect the views of National Research Foundation, Singapore.

\section*{Author contributions}
Y.Z. conceived the project. Y.Z. and X.C. performed the experiments. Y.Z. conducted the numerical simulations. Y.Z. and X.C. analyzed and interpreted the results. Y.Z. wrote the manuscript with input from X.C. All authors have read and approved the manuscript.

\section*{Competing interests}
The authors declare no competing interests.

\section*{Supplementary information}
The relevant additional data can be found in the attached Supplementary Information.

\bibliographystyle{elsarticle-num-names} 
\bibliography{bio}

\begin{thebibliography}{72}
\expandafter\ifx\csname natexlab\endcsname\relax\def\natexlab#1{#1}\fi
\providecommand{\url}[1]{\texttt{#1}}
\providecommand{\href}[2]{#2}
\providecommand{\path}[1]{#1}
\providecommand{\DOIprefix}{doi:}
\providecommand{\ArXivprefix}{arXiv:}
\providecommand{\URLprefix}{URL: }
\providecommand{\Pubmedprefix}{pmid:}
\providecommand{\doi}[1]{\href{http://dx.doi.org/#1}{\path{#1}}}
\providecommand{\Pubmed}[1]{\href{pmid:#1}{\path{#1}}}
\providecommand{\bibinfo}[2]{#2}
\ifx\xfnm\relax \def\xfnm[#1]{\unskip,\space#1}\fi
\bibitem[{Harris and Piersol(2002)}]{harris2002harris}
\bibinfo{author}{C.~M. Harris}, \bibinfo{author}{A.~G. Piersol}, \bibinfo{title}{Harris' shock and vibration handbook}, volume~\bibinfo{volume}{5}, \bibinfo{publisher}{McGraw-Hill New York}, \bibinfo{year}{2002}.
\bibitem[{Heertjes et~al.(2020)Heertjes, Butler, Dirkx, Van Der~Meulen, Ahlawat, O’Brien, Simonelli, Teng, and Zhao}]{heertjes2020control}
\bibinfo{author}{M.~F. Heertjes}, \bibinfo{author}{H.~Butler}, \bibinfo{author}{N.~Dirkx}, \bibinfo{author}{S.~Van Der~Meulen}, \bibinfo{author}{R.~Ahlawat}, \bibinfo{author}{K.~O’Brien}, \bibinfo{author}{J.~Simonelli}, \bibinfo{author}{K.~Teng}, \bibinfo{author}{Y.~Zhao},
\newblock \bibinfo{title}{Control of wafer scanners: Methods and developments},
\newblock in: \bibinfo{booktitle}{2020 American Control Conference (ACC)}, \bibinfo{organization}{IEEE}, \bibinfo{year}{2020}, pp. \bibinfo{pages}{3686--3703}.
\bibitem[{Yoon et~al.(2010)Yoon, Lee, Perkins, and Najafi}]{yoon2010analysis}
\bibinfo{author}{S.~W. Yoon}, \bibinfo{author}{S.~Lee}, \bibinfo{author}{N.~C. Perkins}, \bibinfo{author}{K.~Najafi},
\newblock \bibinfo{title}{Analysis and wafer-level design of a high-order silicon vibration isolator for resonating mems devices},
\newblock \bibinfo{journal}{Journal of Micromechanics and Microengineering} \bibinfo{volume}{21} (\bibinfo{year}{2010}) \bibinfo{pages}{015017}.
\bibitem[{Xu et~al.(2016)Xu, Xu, and Chen}]{xu2016intelligent}
\bibinfo{author}{Z.-D. Xu}, \bibinfo{author}{F.-H. Xu}, \bibinfo{author}{X.~Chen},
\newblock \bibinfo{title}{Intelligent vibration isolation and mitigation of a platform by using mr and ve devices},
\newblock \bibinfo{journal}{Journal of Aerospace Engineering} \bibinfo{volume}{29} (\bibinfo{year}{2016}) \bibinfo{pages}{04016010}.
\bibitem[{Liu et~al.(2015)Liu, Jing, Daley, and Li}]{liu2015recent}
\bibinfo{author}{C.~Liu}, \bibinfo{author}{X.~Jing}, \bibinfo{author}{S.~Daley}, \bibinfo{author}{F.~Li},
\newblock \bibinfo{title}{Recent advances in micro-vibration isolation},
\newblock \bibinfo{journal}{Mechanical Systems and Signal Processing} \bibinfo{volume}{56} (\bibinfo{year}{2015}) \bibinfo{pages}{55--80}.
\bibitem[{Chapain and Aly(2019)}]{chapain2019vibration}
\bibinfo{author}{S.~Chapain}, \bibinfo{author}{A.~M. Aly},
\newblock \bibinfo{title}{Vibration attenuation in high-rise buildings to achieve system-level performance under multiple hazards},
\newblock \bibinfo{journal}{Engineering Structures} \bibinfo{volume}{197} (\bibinfo{year}{2019}) \bibinfo{pages}{109352}.
\bibitem[{Ismail(2015)}]{ismail2015inner}
\bibinfo{author}{M.~Ismail},
\newblock \bibinfo{title}{Inner pounding control of the rnc isolator and its impact on seismic isolation efficiency under near-fault earthquakes},
\newblock \bibinfo{journal}{Engineering Structures} \bibinfo{volume}{86} (\bibinfo{year}{2015}) \bibinfo{pages}{99--121}.
\bibitem[{Accadia et~al.(2011)Accadia, Acernese, Antonucci, Astone, Ballardin, Barone, Barsuglia, Bauer, Beker, Belletoile et~al.}]{accadia2011seismic}
\bibinfo{author}{T.~Accadia}, \bibinfo{author}{F.~Acernese}, \bibinfo{author}{F.~Antonucci}, \bibinfo{author}{P.~Astone}, \bibinfo{author}{G.~Ballardin}, \bibinfo{author}{F.~Barone}, \bibinfo{author}{M.~Barsuglia}, \bibinfo{author}{T.~S. Bauer}, \bibinfo{author}{M.~Beker}, \bibinfo{author}{A.~Belletoile}, et~al.,
\newblock \bibinfo{title}{The seismic superattenuators of the virgo gravitational waves interferometer},
\newblock \bibinfo{journal}{Journal of low frequency noise, vibration and active control} \bibinfo{volume}{30} (\bibinfo{year}{2011}) \bibinfo{pages}{63--79}.
\bibitem[{Braccini et~al.(2005)Braccini, Barsotti, Bradaschia, Cella, Di~Virgilio, Ferrante, Fidecaro, Fiori, Frasconi, Gennai et~al.}]{braccini2005measurement}
\bibinfo{author}{S.~Braccini}, \bibinfo{author}{L.~Barsotti}, \bibinfo{author}{C.~Bradaschia}, \bibinfo{author}{G.~Cella}, \bibinfo{author}{A.~Di~Virgilio}, \bibinfo{author}{I.~Ferrante}, \bibinfo{author}{F.~Fidecaro}, \bibinfo{author}{I.~Fiori}, \bibinfo{author}{F.~Frasconi}, \bibinfo{author}{A.~Gennai}, et~al.,
\newblock \bibinfo{title}{Measurement of the seismic attenuation performance of the virgo superattenuator},
\newblock \bibinfo{journal}{Astroparticle Physics} \bibinfo{volume}{23} (\bibinfo{year}{2005}) \bibinfo{pages}{557--565}.
\bibitem[{Matichard et~al.(2015)Matichard, Lantz, Mittleman, Mason, Kissel, Abbott, Biscans, McIver, Abbott, Abbott et~al.}]{matichard2015seismic}
\bibinfo{author}{F.~Matichard}, \bibinfo{author}{B.~Lantz}, \bibinfo{author}{R.~Mittleman}, \bibinfo{author}{K.~Mason}, \bibinfo{author}{J.~Kissel}, \bibinfo{author}{B.~Abbott}, \bibinfo{author}{S.~Biscans}, \bibinfo{author}{J.~McIver}, \bibinfo{author}{R.~Abbott}, \bibinfo{author}{S.~Abbott}, et~al.,
\newblock \bibinfo{title}{Seismic isolation of advanced ligo: Review of strategy, instrumentation and performance},
\newblock \bibinfo{journal}{Classical and Quantum Gravity} \bibinfo{volume}{32} (\bibinfo{year}{2015}) \bibinfo{pages}{185003}.
\bibitem[{Yun and Li(2011)}]{yun2011general}
\bibinfo{author}{Y.~Yun}, \bibinfo{author}{Y.~Li},
\newblock \bibinfo{title}{A general dynamics and control model of a class of multi-dof manipulators for active vibration control},
\newblock \bibinfo{journal}{Mechanism and machine theory} \bibinfo{volume}{46} (\bibinfo{year}{2011}) \bibinfo{pages}{1549--1574}.
\bibitem[{Xianmin et~al.(2002)Xianmin, Changjian, and Erdman}]{xianmin2002active}
\bibinfo{author}{Z.~Xianmin}, \bibinfo{author}{S.~Changjian}, \bibinfo{author}{A.~G. Erdman},
\newblock \bibinfo{title}{Active vibration controller design and comparison study of flexible linkage mechanism systems},
\newblock \bibinfo{journal}{Mechanism and Machine Theory} \bibinfo{volume}{37} (\bibinfo{year}{2002}) \bibinfo{pages}{985--997}.
\bibitem[{Li et~al.(2020)Li, Li, and Li}]{li2020negative}
\bibinfo{author}{H.~Li}, \bibinfo{author}{Y.~Li}, \bibinfo{author}{J.~Li},
\newblock \bibinfo{title}{Negative stiffness devices for vibration isolation applications: a review},
\newblock \bibinfo{journal}{Advances in Structural Engineering} \bibinfo{volume}{23} (\bibinfo{year}{2020}) \bibinfo{pages}{1739--1755}.
\bibitem[{Narimani et~al.(2004)Narimani, Golnaraghi, and Jazar}]{narimani2004frequency}
\bibinfo{author}{A.~Narimani}, \bibinfo{author}{M.~Golnaraghi}, \bibinfo{author}{G.~N. Jazar},
\newblock \bibinfo{title}{Frequency response of a piecewise linear vibration isolator},
\newblock \bibinfo{journal}{Journal of Vibration and control} \bibinfo{volume}{10} (\bibinfo{year}{2004}) \bibinfo{pages}{1775--1794}.
\bibitem[{Yang et~al.(2021)Yang, Cao, and Hao}]{yang2021novel}
\bibinfo{author}{T.~Yang}, \bibinfo{author}{Q.~Cao}, \bibinfo{author}{Z.~Hao},
\newblock \bibinfo{title}{A novel nonlinear mechanical oscillator and its application in vibration isolation and energy harvesting},
\newblock \bibinfo{journal}{Mechanical systems and signal processing} \bibinfo{volume}{155} (\bibinfo{year}{2021}) \bibinfo{pages}{107636}.
\bibitem[{Liu et~al.(2024)Liu, Zhang, Yu, Liu, and Zheng}]{liu2024quasi}
\bibinfo{author}{C.~Liu}, \bibinfo{author}{W.~Zhang}, \bibinfo{author}{K.~Yu}, \bibinfo{author}{T.~Liu}, \bibinfo{author}{Y.~Zheng},
\newblock \bibinfo{title}{Quasi-zero-stiffness vibration isolation: Designs, improvements and applications},
\newblock \bibinfo{journal}{Engineering Structures} \bibinfo{volume}{301} (\bibinfo{year}{2024}) \bibinfo{pages}{117282}.
\bibitem[{Jing et~al.(2022)Jing, Chai, Chao, and Bian}]{jing2022situ}
\bibinfo{author}{X.~Jing}, \bibinfo{author}{Y.~Chai}, \bibinfo{author}{X.~Chao}, \bibinfo{author}{J.~Bian},
\newblock \bibinfo{title}{In-situ adjustable nonlinear passive stiffness using x-shaped mechanisms},
\newblock \bibinfo{journal}{Mechanical Systems and Signal Processing} \bibinfo{volume}{170} (\bibinfo{year}{2022}) \bibinfo{pages}{108267}.
\bibitem[{Carrella et~al.(2007)Carrella, Brennan, and Waters}]{carrella2007static}
\bibinfo{author}{A.~Carrella}, \bibinfo{author}{M.~Brennan}, \bibinfo{author}{T.~Waters},
\newblock \bibinfo{title}{Static analysis of a passive vibration isolator with quasi-zero-stiffness characteristic},
\newblock \bibinfo{journal}{Journal of sound and vibration} \bibinfo{volume}{301} (\bibinfo{year}{2007}) \bibinfo{pages}{678--689}.
\bibitem[{Carrella et~al.(2009)Carrella, Brennan, Kovacic, and Waters}]{carrella2009force}
\bibinfo{author}{A.~Carrella}, \bibinfo{author}{M.~Brennan}, \bibinfo{author}{I.~Kovacic}, \bibinfo{author}{T.~Waters},
\newblock \bibinfo{title}{On the force transmissibility of a vibration isolator with quasi-zero-stiffness},
\newblock \bibinfo{journal}{Journal of Sound and Vibration} \bibinfo{volume}{322} (\bibinfo{year}{2009}) \bibinfo{pages}{707--717}.
\bibitem[{Huang et~al.(2014)Huang, Liu, Sun, Zhang, and Hua}]{huang2014vibration}
\bibinfo{author}{X.~Huang}, \bibinfo{author}{X.~Liu}, \bibinfo{author}{J.~Sun}, \bibinfo{author}{Z.~Zhang}, \bibinfo{author}{H.~Hua},
\newblock \bibinfo{title}{Vibration isolation characteristics of a nonlinear isolator using euler buckled beam as negative stiffness corrector: a theoretical and experimental study},
\newblock \bibinfo{journal}{Journal of Sound and Vibration} \bibinfo{volume}{333} (\bibinfo{year}{2014}) \bibinfo{pages}{1132--1148}.
\bibitem[{Robertson et~al.(2009)Robertson, Kidner, Cazzolato, and Zander}]{robertson2009theoretical}
\bibinfo{author}{W.~S. Robertson}, \bibinfo{author}{M.~Kidner}, \bibinfo{author}{B.~S. Cazzolato}, \bibinfo{author}{A.~C. Zander},
\newblock \bibinfo{title}{Theoretical design parameters for a quasi-zero stiffness magnetic spring for vibration isolation},
\newblock \bibinfo{journal}{Journal of Sound and Vibration} \bibinfo{volume}{326} (\bibinfo{year}{2009}) \bibinfo{pages}{88--103}.
\bibitem[{Jiang et~al.(2020)Jiang, Song, Ding, and Xu}]{jiang2020design}
\bibinfo{author}{Y.~Jiang}, \bibinfo{author}{C.~Song}, \bibinfo{author}{C.~Ding}, \bibinfo{author}{B.~Xu},
\newblock \bibinfo{title}{Design of magnetic-air hybrid quasi-zero stiffness vibration isolation system},
\newblock \bibinfo{journal}{Journal of Sound and Vibration} \bibinfo{volume}{477} (\bibinfo{year}{2020}) \bibinfo{pages}{115346}.
\bibitem[{Carrella et~al.(2007)Carrella, Brennan, and Waters}]{carrella2007optimization}
\bibinfo{author}{A.~Carrella}, \bibinfo{author}{M.~Brennan}, \bibinfo{author}{T.~Waters},
\newblock \bibinfo{title}{Optimization of a quasi-zero-stiffness isolator},
\newblock \bibinfo{journal}{Journal of Mechanical Science and Technology} \bibinfo{volume}{21} (\bibinfo{year}{2007}) \bibinfo{pages}{946--949}.
\bibitem[{Yan et~al.(2019)Yan, Ma, Jian, Wang, and Wu}]{yan2019nonlinear}
\bibinfo{author}{B.~Yan}, \bibinfo{author}{H.~Ma}, \bibinfo{author}{B.~Jian}, \bibinfo{author}{K.~Wang}, \bibinfo{author}{C.~Wu},
\newblock \bibinfo{title}{Nonlinear dynamics analysis of a bi-state nonlinear vibration isolator with symmetric permanent magnets},
\newblock \bibinfo{journal}{Nonlinear Dynamics} \bibinfo{volume}{97} (\bibinfo{year}{2019}) \bibinfo{pages}{2499--2519}.
\bibitem[{Yan et~al.(2018)Yan, Ma, Zhao, Wu, Wang, and Wang}]{yan2018vari}
\bibinfo{author}{B.~Yan}, \bibinfo{author}{H.~Ma}, \bibinfo{author}{C.~Zhao}, \bibinfo{author}{C.~Wu}, \bibinfo{author}{K.~Wang}, \bibinfo{author}{P.~Wang},
\newblock \bibinfo{title}{A vari-stiffness nonlinear isolator with magnetic effects: Theoretical modeling and experimental verification},
\newblock \bibinfo{journal}{International Journal of Mechanical Sciences} \bibinfo{volume}{148} (\bibinfo{year}{2018}) \bibinfo{pages}{745--755}.
\bibitem[{Yan et~al.(2022)Yan, Yu, Wang, Wu, Wang, and Zhang}]{yan2022lever}
\bibinfo{author}{B.~Yan}, \bibinfo{author}{N.~Yu}, \bibinfo{author}{Z.~Wang}, \bibinfo{author}{C.~Wu}, \bibinfo{author}{S.~Wang}, \bibinfo{author}{W.~Zhang},
\newblock \bibinfo{title}{Lever-type quasi-zero stiffness vibration isolator with magnetic spring},
\newblock \bibinfo{journal}{Journal of Sound and Vibration} \bibinfo{volume}{527} (\bibinfo{year}{2022}) \bibinfo{pages}{116865}.
\bibitem[{Liu et~al.(2024)Liu, Wang, Zhang, Yu, Mao, and Shen}]{liu2024nonlinear}
\bibinfo{author}{C.~Liu}, \bibinfo{author}{Y.~Wang}, \bibinfo{author}{W.~Zhang}, \bibinfo{author}{K.~Yu}, \bibinfo{author}{J.-J. Mao}, \bibinfo{author}{H.~Shen},
\newblock \bibinfo{title}{Nonlinear dynamics of a magnetic vibration isolator with higher-order stable quasi-zero-stiffness},
\newblock \bibinfo{journal}{Mechanical Systems and Signal Processing} \bibinfo{volume}{218} (\bibinfo{year}{2024}) \bibinfo{pages}{111584}.
\bibitem[{Yan et~al.(2022)Yan, Zou, Wang, Zhao, Wu, and Zhang}]{yan2022bio}
\bibinfo{author}{G.~Yan}, \bibinfo{author}{H.-X. Zou}, \bibinfo{author}{S.~Wang}, \bibinfo{author}{L.-C. Zhao}, \bibinfo{author}{Z.-Y. Wu}, \bibinfo{author}{W.-M. Zhang},
\newblock \bibinfo{title}{Bio-inspired toe-like structure for low-frequency vibration isolation},
\newblock \bibinfo{journal}{Mechanical Systems and Signal Processing} \bibinfo{volume}{162} (\bibinfo{year}{2022}) \bibinfo{pages}{108010}.
\bibitem[{Jing et~al.(2019)Jing, Zhang, Feng, Sun, and Li}]{jing2019novel}
\bibinfo{author}{X.~Jing}, \bibinfo{author}{L.~Zhang}, \bibinfo{author}{X.~Feng}, \bibinfo{author}{B.~Sun}, \bibinfo{author}{Q.~Li},
\newblock \bibinfo{title}{A novel bio-inspired anti-vibration structure for operating hand-held jackhammers},
\newblock \bibinfo{journal}{Mechanical systems and signal processing} \bibinfo{volume}{118} (\bibinfo{year}{2019}) \bibinfo{pages}{317--339}.
\bibitem[{Wu et~al.(2020)Wu, Wang, Zhai, Yang, Krishnaraju, Lu, Wu, Wang, and Jiang}]{wu2020mechanical}
\bibinfo{author}{L.~Wu}, \bibinfo{author}{Y.~Wang}, \bibinfo{author}{Z.~Zhai}, \bibinfo{author}{Y.~Yang}, \bibinfo{author}{D.~Krishnaraju}, \bibinfo{author}{J.~Lu}, \bibinfo{author}{F.~Wu}, \bibinfo{author}{Q.~Wang}, \bibinfo{author}{H.~Jiang},
\newblock \bibinfo{title}{Mechanical metamaterials for full-band mechanical wave shielding},
\newblock \bibinfo{journal}{Applied Materials Today} \bibinfo{volume}{20} (\bibinfo{year}{2020}) \bibinfo{pages}{100671}.
\bibitem[{Ishida et~al.(2017)Ishida, Suzuki, and Shimosaka}]{ishida2017design}
\bibinfo{author}{S.~Ishida}, \bibinfo{author}{K.~Suzuki}, \bibinfo{author}{H.~Shimosaka},
\newblock \bibinfo{title}{Design and experimental analysis of origami-inspired vibration isolator with quasi-zero-stiffness characteristic},
\newblock \bibinfo{journal}{Journal of Vibration and Acoustics} \bibinfo{volume}{139} (\bibinfo{year}{2017}) \bibinfo{pages}{051004}.
\bibitem[{Yan et~al.(2020)Yan, Zou, Wang, Zhao, Gao, Tan, and Zhang}]{yan2020large}
\bibinfo{author}{G.~Yan}, \bibinfo{author}{H.-X. Zou}, \bibinfo{author}{S.~Wang}, \bibinfo{author}{L.-C. Zhao}, \bibinfo{author}{Q.-H. Gao}, \bibinfo{author}{T.~Tan}, \bibinfo{author}{W.-M. Zhang},
\newblock \bibinfo{title}{Large stroke quasi-zero stiffness vibration isolator using three-link mechanism},
\newblock \bibinfo{journal}{Journal of Sound and Vibration} \bibinfo{volume}{478} (\bibinfo{year}{2020}) \bibinfo{pages}{115344}.
\bibitem[{Bertoldi et~al.(2017)Bertoldi, Vitelli, Christensen, and Van~Hecke}]{bertoldi2017flexible}
\bibinfo{author}{K.~Bertoldi}, \bibinfo{author}{V.~Vitelli}, \bibinfo{author}{J.~Christensen}, \bibinfo{author}{M.~Van~Hecke},
\newblock \bibinfo{title}{Flexible mechanical metamaterials},
\newblock \bibinfo{journal}{Nature Reviews Materials} \bibinfo{volume}{2} (\bibinfo{year}{2017}) \bibinfo{pages}{1--11}.
\bibitem[{Babaee et~al.(2013)Babaee, Shim, Weaver, Chen, Patel, and Bertoldi}]{babaee20133d}
\bibinfo{author}{S.~Babaee}, \bibinfo{author}{J.~Shim}, \bibinfo{author}{J.~C. Weaver}, \bibinfo{author}{E.~R. Chen}, \bibinfo{author}{N.~Patel}, \bibinfo{author}{K.~Bertoldi},
\newblock \bibinfo{title}{3d soft metamaterials with negative poisson’s ratio},
\newblock \bibinfo{journal}{Adv. Mater} \bibinfo{volume}{25} (\bibinfo{year}{2013}) \bibinfo{pages}{5044--5049}.
\bibitem[{Huang and Chen(2016)}]{huang2016negative}
\bibinfo{author}{C.~Huang}, \bibinfo{author}{L.~Chen},
\newblock \bibinfo{title}{Negative poisson's ratio in modern functional materials},
\newblock \bibinfo{journal}{Advanced Materials} \bibinfo{volume}{28} (\bibinfo{year}{2016}) \bibinfo{pages}{8079--8096}.
\bibitem[{Zhang et~al.(2021)Zhang, Tichem, and van Keulen}]{zhang2021novel}
\bibinfo{author}{Y.~Zhang}, \bibinfo{author}{M.~Tichem}, \bibinfo{author}{F.~van Keulen},
\newblock \bibinfo{title}{A novel design of multi-stable metastructures for energy dissipation},
\newblock \bibinfo{journal}{Materials \& Design} \bibinfo{volume}{212} (\bibinfo{year}{2021}) \bibinfo{pages}{110234}.
\bibitem[{Xiu et~al.(2022)Xiu, Liu, Poli, Wan, Sun, Arruda, Mao, and Chen}]{xiu2022topological}
\bibinfo{author}{H.~Xiu}, \bibinfo{author}{H.~Liu}, \bibinfo{author}{A.~Poli}, \bibinfo{author}{G.~Wan}, \bibinfo{author}{K.~Sun}, \bibinfo{author}{E.~M. Arruda}, \bibinfo{author}{X.~Mao}, \bibinfo{author}{Z.~Chen},
\newblock \bibinfo{title}{Topological transformability and reprogrammability of multistable mechanical metamaterials},
\newblock \bibinfo{journal}{Proceedings of the National Academy of Sciences} \bibinfo{volume}{119} (\bibinfo{year}{2022}) \bibinfo{pages}{e2211725119}.
\bibitem[{Zhang et~al.(2020)Zhang, Wang, Tichem, and van Keulen}]{zhang2020design}
\bibinfo{author}{Y.~Zhang}, \bibinfo{author}{Q.~Wang}, \bibinfo{author}{M.~Tichem}, \bibinfo{author}{F.~van Keulen},
\newblock \bibinfo{title}{Design and characterization of multi-stable mechanical metastructures with level and tilted stable configurations},
\newblock \bibinfo{journal}{Extreme Mechanics Letters} \bibinfo{volume}{34} (\bibinfo{year}{2020}) \bibinfo{pages}{100593}.
\bibitem[{Mofatteh et~al.(2022)Mofatteh, Shahryari, Mirabolghasemi, Seyedkanani, Shirzadkhani, Desharnais, and Akbarzadeh}]{mofatteh2022programming}
\bibinfo{author}{H.~Mofatteh}, \bibinfo{author}{B.~Shahryari}, \bibinfo{author}{A.~Mirabolghasemi}, \bibinfo{author}{A.~Seyedkanani}, \bibinfo{author}{R.~Shirzadkhani}, \bibinfo{author}{G.~Desharnais}, \bibinfo{author}{A.~Akbarzadeh},
\newblock \bibinfo{title}{Programming multistable metamaterials to discover latent functionalities},
\newblock \bibinfo{journal}{Advanced Science} \bibinfo{volume}{9} (\bibinfo{year}{2022}) \bibinfo{pages}{2202883}.
\bibitem[{Wu et~al.(2016)Wu, Li, and Zhou}]{wu2016isotropic}
\bibinfo{author}{L.~Wu}, \bibinfo{author}{B.~Li}, \bibinfo{author}{J.~Zhou},
\newblock \bibinfo{title}{Isotropic negative thermal expansion metamaterials},
\newblock \bibinfo{journal}{ACS applied materials \& interfaces} \bibinfo{volume}{8} (\bibinfo{year}{2016}) \bibinfo{pages}{17721--17727}.
\bibitem[{Wang et~al.(2016)Wang, Jackson, Ge, Hopkins, Spadaccini, and Fang}]{wang2016lightweight}
\bibinfo{author}{Q.~Wang}, \bibinfo{author}{J.~A. Jackson}, \bibinfo{author}{Q.~Ge}, \bibinfo{author}{J.~B. Hopkins}, \bibinfo{author}{C.~M. Spadaccini}, \bibinfo{author}{N.~X. Fang},
\newblock \bibinfo{title}{Lightweight mechanical metamaterials with tunable negative thermal expansion},
\newblock \bibinfo{journal}{Physical review letters} \bibinfo{volume}{117} (\bibinfo{year}{2016}) \bibinfo{pages}{175901}.
\bibitem[{Dalela et~al.(2022)Dalela, Balaji, and Jena}]{dalela2022review}
\bibinfo{author}{S.~Dalela}, \bibinfo{author}{P.~Balaji}, \bibinfo{author}{D.~Jena},
\newblock \bibinfo{title}{A review on application of mechanical metamaterials for vibration control},
\newblock \bibinfo{journal}{Mechanics of advanced materials and structures} \bibinfo{volume}{29} (\bibinfo{year}{2022}) \bibinfo{pages}{3237--3262}.
\bibitem[{Ji et~al.(2021)Ji, Luo, and Ye}]{ji2021vibration}
\bibinfo{author}{J.~Ji}, \bibinfo{author}{Q.~Luo}, \bibinfo{author}{K.~Ye},
\newblock \bibinfo{title}{Vibration control based metamaterials and origami structures: A state-of-the-art review},
\newblock \bibinfo{journal}{Mechanical Systems and Signal Processing} \bibinfo{volume}{161} (\bibinfo{year}{2021}) \bibinfo{pages}{107945}.
\bibitem[{Liu et~al.(2025)Liu, Wu, Zhang, Sun, and Zhou}]{liu2025metamaterial}
\bibinfo{author}{W.~Liu}, \bibinfo{author}{L.~Wu}, \bibinfo{author}{J.~Zhang}, \bibinfo{author}{J.~Sun}, \bibinfo{author}{J.~Zhou},
\newblock \bibinfo{title}{Metamaterial springs for low-frequency vibration isolation},
\newblock \bibinfo{journal}{Journal of Materiomics} \bibinfo{volume}{11} (\bibinfo{year}{2025}) \bibinfo{pages}{100884}.
\bibitem[{Song et~al.(2025)Song, Yan, Wang, Zhang, Xue, Liu, Tian, Wu, Jiang, and Li}]{song2025genetic}
\bibinfo{author}{X.~Song}, \bibinfo{author}{S.~Yan}, \bibinfo{author}{Y.~Wang}, \bibinfo{author}{H.~Zhang}, \bibinfo{author}{J.~Xue}, \bibinfo{author}{T.~Liu}, \bibinfo{author}{X.~Tian}, \bibinfo{author}{L.~Wu}, \bibinfo{author}{H.~Jiang}, \bibinfo{author}{D.~Li},
\newblock \bibinfo{title}{Genetic algorithm-enabled mechanical metamaterials for vibration isolation with different payloads},
\newblock \bibinfo{journal}{Journal of Materiomics} \bibinfo{volume}{11} (\bibinfo{year}{2025}) \bibinfo{pages}{100944}.
\bibitem[{Banerjee et~al.(2023)Banerjee, Dalela, Balaji, Murugan, and Kumaraswamidhas}]{banerjee2023simultaneous}
\bibinfo{author}{P.~Banerjee}, \bibinfo{author}{S.~Dalela}, \bibinfo{author}{P.~Balaji}, \bibinfo{author}{S.~Murugan}, \bibinfo{author}{L.~Kumaraswamidhas},
\newblock \bibinfo{title}{Simultaneous vibration isolation and energy harvesting using quasi-zero-stiffness-based metastructure},
\newblock \bibinfo{journal}{Acta Mechanica} \bibinfo{volume}{234} (\bibinfo{year}{2023}) \bibinfo{pages}{3337--3359}.
\bibitem[{Dalela et~al.(2022)Dalela, Balaji, and Jena}]{dalela2022design}
\bibinfo{author}{S.~Dalela}, \bibinfo{author}{P.~Balaji}, \bibinfo{author}{D.~Jena},
\newblock \bibinfo{title}{Design of a metastructure for vibration isolation with quasi-zero-stiffness characteristics using bistable curved beam},
\newblock \bibinfo{journal}{Nonlinear Dynamics} \bibinfo{volume}{108} (\bibinfo{year}{2022}) \bibinfo{pages}{1931--1971}.
\bibitem[{Fan et~al.(2020)Fan, Yang, Tian, and Wang}]{fan2020design}
\bibinfo{author}{H.~Fan}, \bibinfo{author}{L.~Yang}, \bibinfo{author}{Y.~Tian}, \bibinfo{author}{Z.~Wang},
\newblock \bibinfo{title}{Design of metastructures with quasi-zero dynamic stiffness for vibration isolation},
\newblock \bibinfo{journal}{Composite Structures} \bibinfo{volume}{243} (\bibinfo{year}{2020}) \bibinfo{pages}{112244}.
\bibitem[{Cai et~al.(2020)Cai, Zhou, Wu, Wang, Xu, and Ouyang}]{cai2020design}
\bibinfo{author}{C.~Cai}, \bibinfo{author}{J.~Zhou}, \bibinfo{author}{L.~Wu}, \bibinfo{author}{K.~Wang}, \bibinfo{author}{D.~Xu}, \bibinfo{author}{H.~Ouyang},
\newblock \bibinfo{title}{Design and numerical validation of quasi-zero-stiffness metamaterials for very low-frequency band gaps},
\newblock \bibinfo{journal}{Composite structures} \bibinfo{volume}{236} (\bibinfo{year}{2020}) \bibinfo{pages}{111862}.
\bibitem[{Zhang et~al.(2023)Zhang, He, Zhou, Wang, Xu, and Zhou}]{zhang2023compliant}
\bibinfo{author}{C.~Zhang}, \bibinfo{author}{J.~He}, \bibinfo{author}{G.~Zhou}, \bibinfo{author}{K.~Wang}, \bibinfo{author}{D.~Xu}, \bibinfo{author}{J.~Zhou},
\newblock \bibinfo{title}{Compliant quasi-zero-stiffness isolator for low-frequency torsional vibration isolation},
\newblock \bibinfo{journal}{Mechanism and Machine Theory} \bibinfo{volume}{181} (\bibinfo{year}{2023}) \bibinfo{pages}{105213}.
\bibitem[{Yu et~al.(2022)Yu, Fu, Zhang, and Zhang}]{yu2022vibration}
\bibinfo{author}{C.~Yu}, \bibinfo{author}{Q.~Fu}, \bibinfo{author}{J.~Zhang}, \bibinfo{author}{N.~Zhang},
\newblock \bibinfo{title}{The vibration isolation characteristics of torsion bar spring with negative stiffness structure},
\newblock \bibinfo{journal}{Mechanical Systems and Signal Processing} \bibinfo{volume}{180} (\bibinfo{year}{2022}) \bibinfo{pages}{109378}.
\bibitem[{Zhou et~al.(2021)Zhou, Pan, Cai, and Xu}]{zhou2021tunable}
\bibinfo{author}{J.~Zhou}, \bibinfo{author}{H.~Pan}, \bibinfo{author}{C.~Cai}, \bibinfo{author}{D.~Xu},
\newblock \bibinfo{title}{Tunable ultralow frequency wave attenuations in one-dimensional quasi-zero-stiffness metamaterial},
\newblock \bibinfo{journal}{International Journal of Mechanics and Materials in Design} \bibinfo{volume}{17} (\bibinfo{year}{2021}) \bibinfo{pages}{285--300}.
\bibitem[{Dalela et~al.(2024)Dalela, Balaji, Leblouba, Trivedi, and Kalam}]{dalela2024nonlinear}
\bibinfo{author}{S.~Dalela}, \bibinfo{author}{P.~Balaji}, \bibinfo{author}{M.~Leblouba}, \bibinfo{author}{S.~Trivedi}, \bibinfo{author}{A.~Kalam},
\newblock \bibinfo{title}{Nonlinear static and dynamic response of a metastructure exhibiting quasi-zero-stiffness characteristics for vibration control: an experimental validation},
\newblock \bibinfo{journal}{Scientific Reports} \bibinfo{volume}{14} (\bibinfo{year}{2024}) \bibinfo{pages}{19195}.
\bibitem[{Zhang et~al.(2021)Zhang, Guo, and Hu}]{zhang2021tailored}
\bibinfo{author}{Q.~Zhang}, \bibinfo{author}{D.~Guo}, \bibinfo{author}{G.~Hu},
\newblock \bibinfo{title}{Tailored mechanical metamaterials with programmable quasi-zero-stiffness features for full-band vibration isolation},
\newblock \bibinfo{journal}{Advanced Functional Materials} \bibinfo{volume}{31} (\bibinfo{year}{2021}) \bibinfo{pages}{2101428}.
\bibitem[{Hou and Wei(2024)}]{hou2024quasi}
\bibinfo{author}{S.~Hou}, \bibinfo{author}{J.~Wei},
\newblock \bibinfo{title}{A quasi-zero stiffness mechanism with monolithic flexible beams for low-frequency vibration isolation},
\newblock \bibinfo{journal}{Mechanical Systems and Signal Processing} \bibinfo{volume}{210} (\bibinfo{year}{2024}) \bibinfo{pages}{111154}.
\bibitem[{Liu et~al.(2024)Liu, Zhang, Wu, Sun, and Zhou}]{liu2024design}
\bibinfo{author}{W.~Liu}, \bibinfo{author}{Q.~Zhang}, \bibinfo{author}{L.~Wu}, \bibinfo{author}{J.~Sun}, \bibinfo{author}{J.~Zhou},
\newblock \bibinfo{title}{Design of quasi-zero stiffness metamaterials with high reliability via metallic architected materials},
\newblock \bibinfo{journal}{Thin-Walled Structures} \bibinfo{volume}{198} (\bibinfo{year}{2024}) \bibinfo{pages}{111686}.
\bibitem[{Parisch(1978)}]{parisch1978geometrical}
\bibinfo{author}{H.~Parisch},
\newblock \bibinfo{title}{Geometrical nonlinear analysis of shells},
\newblock \bibinfo{journal}{Computer Methods in Applied Mechanics and Engineering} \bibinfo{volume}{14} (\bibinfo{year}{1978}) \bibinfo{pages}{159--178}.
\bibitem[{Liu et~al.(2024)Liu, Chen, Wang, Tan, and Yu}]{liu2024compact}
\bibinfo{author}{X.~Liu}, \bibinfo{author}{S.~Chen}, \bibinfo{author}{B.~Wang}, \bibinfo{author}{X.~Tan}, \bibinfo{author}{L.~Yu},
\newblock \bibinfo{title}{A compact quasi-zero-stiffness mechanical metamaterial based on truncated conical shells},
\newblock \bibinfo{journal}{International Journal of Mechanical Sciences} \bibinfo{volume}{277} (\bibinfo{year}{2024}) \bibinfo{pages}{109390}.
\bibitem[{Qiu et~al.(2004)Qiu, Lang, and Slocum}]{qiu2004curved}
\bibinfo{author}{J.~Qiu}, \bibinfo{author}{J.~H. Lang}, \bibinfo{author}{A.~H. Slocum},
\newblock \bibinfo{title}{A curved-beam bistable mechanism},
\newblock \bibinfo{journal}{Journal of microelectromechanical systems} \bibinfo{volume}{13} (\bibinfo{year}{2004}) \bibinfo{pages}{137--146}.
\bibitem[{Dalela et~al.(2024)Dalela, Prasad, Balaji, Trivedi, and Kalam}]{dalela2024novel}
\bibinfo{author}{S.~Dalela}, \bibinfo{author}{P.~Prasad}, \bibinfo{author}{P.~Balaji}, \bibinfo{author}{S.~Trivedi}, \bibinfo{author}{A.~Kalam},
\newblock \bibinfo{title}{A novel bottom reinforced cosine beam based metastructure with quasi zero stiffness characteristics for vibration isolation applications},
\newblock in: \bibinfo{booktitle}{Structures}, volume~\bibinfo{volume}{67}, \bibinfo{organization}{Elsevier}, \bibinfo{year}{2024}, p. \bibinfo{pages}{106950}.
\bibitem[{Liu et~al.(2025)Liu, Lv, Li, Huan, and Huang}]{liu2025straight}
\bibinfo{author}{Z.~Liu}, \bibinfo{author}{Q.~Lv}, \bibinfo{author}{D.~Li}, \bibinfo{author}{R.~Huan}, \bibinfo{author}{Z.~Huang},
\newblock \bibinfo{title}{A straight-arch-straight beam tandem quasi-zero stiffness structure},
\newblock \bibinfo{journal}{International Journal of Mechanical Sciences} \bibinfo{volume}{286} (\bibinfo{year}{2025}) \bibinfo{pages}{109818}.
\bibitem[{Zhang et~al.(2024)Zhang, Lu, Li, Tian, Chen, and Wang}]{zhang2024design}
\bibinfo{author}{X.~Zhang}, \bibinfo{author}{X.~Lu}, \bibinfo{author}{C.~Li}, \bibinfo{author}{R.~Tian}, \bibinfo{author}{L.~Chen}, \bibinfo{author}{M.~Wang},
\newblock \bibinfo{title}{Design of hyperbolic quasi-zero stiffness metastructures coupled with nonlinear stiffness for low-frequency vibration isolation},
\newblock \bibinfo{journal}{Engineering Structures} \bibinfo{volume}{312} (\bibinfo{year}{2024}) \bibinfo{pages}{118262}.
\bibitem[{Zhou et~al.(2024)Zhou, Sui, Chen, and Shan}]{zhou2024nonlinear}
\bibinfo{author}{C.~Zhou}, \bibinfo{author}{G.~Sui}, \bibinfo{author}{Y.~Chen}, \bibinfo{author}{X.~Shan},
\newblock \bibinfo{title}{A nonlinear low frequency quasi zero stiffness vibration isolator using double-arc flexible beams},
\newblock \bibinfo{journal}{International Journal of Mechanical Sciences} \bibinfo{volume}{276} (\bibinfo{year}{2024}) \bibinfo{pages}{109378}.
\bibitem[{Zhang et~al.(2023)Zhang, Tian, and Xu}]{zhang2023bistable}
\bibinfo{author}{Z.~Zhang}, \bibinfo{author}{J.~Tian}, \bibinfo{author}{Z.-D. Xu},
\newblock \bibinfo{title}{Bistable inclined beam connected in series for quasi-zero stiffness},
\newblock \bibinfo{journal}{Mechanics of Advanced Materials and Structures} \bibinfo{volume}{30} (\bibinfo{year}{2023}) \bibinfo{pages}{1285--1298}.
\bibitem[{Cai et~al.(2022)Cai, Zhou, Wang, Pan, Tan, Xu, and Wen}]{cai2022flexural}
\bibinfo{author}{C.~Cai}, \bibinfo{author}{J.~Zhou}, \bibinfo{author}{K.~Wang}, \bibinfo{author}{H.~Pan}, \bibinfo{author}{D.~Tan}, \bibinfo{author}{D.~Xu}, \bibinfo{author}{G.~Wen},
\newblock \bibinfo{title}{Flexural wave attenuation by metamaterial beam with compliant quasi-zero-stiffness resonators},
\newblock \bibinfo{journal}{Mechanical Systems and Signal Processing} \bibinfo{volume}{174} (\bibinfo{year}{2022}) \bibinfo{pages}{109119}.
\bibitem[{Liang et~al.(2024)Liang, Jing, and Zhang}]{liang2024design}
\bibinfo{author}{K.~Liang}, \bibinfo{author}{Y.~Jing}, \bibinfo{author}{X.~Zhang},
\newblock \bibinfo{title}{Design of broad quasi-zero stiffness platform metamaterials for vibration isolation},
\newblock \bibinfo{journal}{International Journal of Mechanical Sciences} \bibinfo{volume}{281} (\bibinfo{year}{2024}) \bibinfo{pages}{109691}.
\bibitem[{Medel et~al.(2022)Medel, Abad, and Esteban}]{medel2022stiffness}
\bibinfo{author}{F.~Medel}, \bibinfo{author}{J.~Abad}, \bibinfo{author}{V.~Esteban},
\newblock \bibinfo{title}{Stiffness and damping behavior of 3d printed specimens},
\newblock \bibinfo{journal}{Polymer Testing} \bibinfo{volume}{109} (\bibinfo{year}{2022}) \bibinfo{pages}{107529}.
\bibitem[{Zu et~al.(2025)Zu, Chen, Huang, Chen, Du, Fan, Li, and Liu}]{zu2025tailor}
\bibinfo{author}{R.~Zu}, \bibinfo{author}{W.~Chen}, \bibinfo{author}{Y.~Huang}, \bibinfo{author}{Y.~Chen}, \bibinfo{author}{C.~Du}, \bibinfo{author}{Q.~Fan}, \bibinfo{author}{H.~Li}, \bibinfo{author}{H.~Liu},
\newblock \bibinfo{title}{Tailor-made 3d printing tpu/pla composites for damping and energy absorption},
\newblock \bibinfo{journal}{Materials \& Design} \bibinfo{volume}{252} (\bibinfo{year}{2025}) \bibinfo{pages}{113752}.
\bibitem[{Chen et~al.(2024)Chen, de~Lint, Alijani, and Steeneken}]{chen2024nonlinear}
\bibinfo{author}{X.~Chen}, \bibinfo{author}{T.~de~Lint}, \bibinfo{author}{F.~Alijani}, \bibinfo{author}{P.~G. Steeneken},
\newblock \bibinfo{title}{Nonlinear dynamics of diamagnetically levitating resonators},
\newblock \bibinfo{journal}{Nonlinear Dynamics} \bibinfo{volume}{112} (\bibinfo{year}{2024}) \bibinfo{pages}{18807--18816}.
\bibitem[{Bao and Yang(2007)}]{bao2007squeeze}
\bibinfo{author}{M.~Bao}, \bibinfo{author}{H.~Yang},
\newblock \bibinfo{title}{Squeeze film air damping in mems},
\newblock \bibinfo{journal}{Sensors and Actuators A: Physical} \bibinfo{volume}{136} (\bibinfo{year}{2007}) \bibinfo{pages}{3--27}.
\bibitem[{Neubauer et~al.(2022)Neubauer, Pohl, Kucher, B{\"o}hm, H{\"o}schler, and Modler}]{neubauer2022dma}
\bibinfo{author}{M.~Neubauer}, \bibinfo{author}{M.~Pohl}, \bibinfo{author}{M.~Kucher}, \bibinfo{author}{R.~B{\"o}hm}, \bibinfo{author}{K.~H{\"o}schler}, \bibinfo{author}{N.~Modler},
\newblock \bibinfo{title}{Dma of tpu films and the modelling of their viscoelastic properties for noise reduction in jet engines},
\newblock \bibinfo{journal}{Polymers} \bibinfo{volume}{14} (\bibinfo{year}{2022}) \bibinfo{pages}{5285}.
\bibitem[{Shan et~al.(2015)Shan, Kang, Raney, Wang, Fang, Candido, Lewis, and Bertoldi}]{shan2015multistable}
\bibinfo{author}{S.~Shan}, \bibinfo{author}{S.~H. Kang}, \bibinfo{author}{J.~R. Raney}, \bibinfo{author}{P.~Wang}, \bibinfo{author}{L.~Fang}, \bibinfo{author}{F.~Candido}, \bibinfo{author}{J.~A. Lewis}, \bibinfo{author}{K.~Bertoldi},
\newblock \bibinfo{title}{Multistable architected materials for trapping elastic strain energy},
\newblock \bibinfo{journal}{Advanced Materials} \bibinfo{volume}{27} (\bibinfo{year}{2015}) \bibinfo{pages}{4296--4301}.

\end{thebibliography}

\end{document}